\documentclass[preprint,showpacs,preprintnumbers,amsmath,amssymb]{revtex4}

\usepackage{epsfig}
\usepackage{graphicx}
\usepackage{dcolumn}
\usepackage{bm}
\def\la{\hbox{{\lower -2.5pt\hbox{$<$}}\hskip -8pt\raise
-2.5pt\hbox{$\sim$}}}
\def\ga{\hbox{{\lower -2.5pt\hbox{$>$}}\hskip -8pt\raise
-2.5pt\hbox{$\sim$}}}

\newcommand{\rb}[1]{\raisebox{1.5ex}{#1}}

\begin{document}
\renewcommand{\thefootnote}{\fnsymbol{footnote}}
\title{The Detectability of Neutralino Clumps via
Atmospheric Cherenkov Telescopes}
\author{Argyro Tasitsiomi\footnote[2]{iro@oddjob.uchicago.edu}  and
Angela V. Olinto\footnote[3]{olinto@oddjob.uchicago.edu}}
\affiliation{Department of Astronomy \& Astrophysics, \\ Enrico Fermi
Institute, \\ Center for Cosmological Physics,  \\  The University of
Chicago,\\ 5640 S. Ellis Ave., Chicago, IL 60637, USA}

\begin{abstract}
High resolution N-body simulations have revealed the survival of
considerable substructure within galactic halos. Assuming that
the predicted dark matter clumps are
composed of annihilating neutralinos, we examine their
detectability via Atmospheric Cherenkov Telescopes
(ACTs). Depending on their density profile, individual
neutralino clumps should be observable via their $\gamma$-ray
continuum and line emissions. We find that the continuum signal is the
most promising signal for detecting a neutralino clump, being
significantly stronger than the monochromatic  signals.  Limits from
the line detectability can help lift degeneracies in the supersymmetric (SUSY)
parameter space.  We show that by combining the observations of different
mass clumps, ACTs can explore most of the SUSY
parameter space. ACTs can play a complementary role to accelerator
and $\gamma$-ray satellite limits
by exploring relatively large  neutralino masses and less
concentrated clumps.
We develop a strategy for dark matter clump studies by future ACTs based on
VERITAS specifications and encourage the development of techniques to identify
primaries. This can reduce the background by an order of magnitude.

\end{abstract}


\maketitle
\section{Introduction}
The presence of a dark matter component is inferred through
gravitational interactions in galaxies and clusters of galaxies.
Its contribution is
estimated to be about 30\% of the critical density of the
Universe. This dark matter cannot be all baryonic. Constraints from
primordial nucleosynthesis and cosmic background radiation measurements
limit the baryonic content of the Universe to be at most about 4\% of
the critical density. The remaining $\simeq$ 26\% of the critical density
is believed to be composed of a, yet to be observed, Cold Dark Matter
(CDM) particle.

Among the several candidates proposed for the non-baryonic dark matter,
the leading scenario involves Weakly Interacting Massive Particles
(WIMPs). Weakly interacting relics from the early universe with masses
from some tens of  GeV to some  TeV can
naturally  give rise to relic densities in
the range of the observed dark matter density. WIMPs are also well
motivated by theoretical extensions of the standard model of
particle physics. In particular, the lightest supersymmetric particle  (LSP)
in supersymmetric extensions of the standard model can be a WIMP
that is stable due to R-parity conservation.
In most scenarios the LSP is neutral and is called the neutralino,
$\chi$ (for a review see, e.g.,~\cite{jkg96}).

Neutralinos may be detected directly as they traverse the Earth, may be
generated in future particle colliders, and may be detected indirectly
by observing  their annihilation products.  Direct neutralino
searches are now underway in a number of low background experiments
with no consensus detection to date. Collider
experiments have been able to place some constraints on the
neutralino parameter space  (see, e.g.,~\cite{efgo00}), but a large region
of the parameter space is still unexplored.
Indirect searches offer a viable
and complementary alternative to direct searches and accelerator
experiments. In particular, neutralino annihilation in the galactic
halo
can be observed via $\gamma$-rays, neutrinos, and synchrotron emission from
the charged annihilation products~\cite{bgz92,bbm94,gs99,bss01,
g00,begu99,cm01,abo02,bot02,t02}.

    The rate of neutralino annihilation is proportional to the neutralino
number density squared ($\sim n^2$), thus the strongest signals are
likely to come from either the galactic center~\cite{bgz92,bbm94,gs99,bss01,
g00,begu99}, or the higher density nearby
clumps of dark matter in our halo~\cite{begu99,cm01,bot02,abo02}. The
dark matter density in the galactic center  depends strongly on the
    formation history of the central black hole~\cite{mm02}, and
its $\gamma$-ray emission due to neutralino annihilation may be
difficult to distinguish from other baryonic $\gamma$-ray sources.
As an alternative, the clumpy nature of CDM halos provides a
number of  high density regions, other than the galactic
center.

Clumps of dark matter distributed throughout galaxy halos is a
generic prediction of high-resolution CDM
simulations (see, e.g.,~\cite{gmglqs,Moore1,Klypin1,iro02}).
This high degree of substructure is what  has remained from
the merging of small halos to form the larger halos (either
cluster or galaxy halos). Simulations show
that about  $10\%$ to $15\%$ of the total mass of a given halo is
in the form of smaller mass clumps. These clumps are ideal sites for
neutralino annihilation and can be observed via
their $\gamma$-ray signal~\cite{begu99,cm01,abo02}, and via
their synchrotron emission~\cite{bot02}. Here,
we study in detail the $\gamma$-ray emission from the
nearest clumps of dark matter via Atmospheric Cherenkov Telescopes
(ACTs), such as VERITAS \cite{veritas}, HESS \cite{hess}, MAGIC
\cite{magic}, and CANGAROO-III \cite{cang3}. We explore the neutralino
parameter space for the continuum,
the $\gamma\gamma$, and the $Z\gamma$ lines and calculate the flux of
different mass clumps assuming different clump density profiles and
VERITAS specifications.
We find that the larger mass clumps can be easily
detected by the next generation ACTs and that the detectability of
the smaller mass
clumps can help constrain the neutralino parameter space. Future ACT
observations are only limited by their threshold energy and
will be more
effective for neutralino masses above
$\sim$ 50 to 100 GeV, a range of neutralino
masses that nicely complements accelerator and satellite
studies of the neutralino parameter space.

In the next section, we discuss the structure and the mass spectrum of
the dark matter clumps that we assume in our study.  In \S III, we present
the neutralino parameter space for continuum $\gamma$-rays,
the $\gamma\gamma$, and the $Z\gamma$ line
emission using DarkSUSY~\cite{darksusy1}. In \S IV, we discuss
the characteristics of ACTs and
the relevant backgrounds. Our results and strategies for CDM clump
detection are shown in \S V. Finally, we conclude in \S VI.

\section{The dark matter clumps}

\subsection{The nearest clumps}

According to N-body simulations~\cite{gmglqs,Moore1,Klypin1},
substructures of masses $\geq 10^{7} M_{\odot}$ survive in significant
numbers in galactic halos. Within the virial radius of a galactic halo
about 10\%-15\% of the  mass is in the form of dark matter clumps. The
mass spectrum for the simulated dark matter clumps in galactic halos is
of the form (e.g.,~\cite{Moore1,Klypin1})
\begin{equation}
\label{distribution}
\frac{dN_{clump}}{dM_{clump}} \propto M_{clump}^{-\alpha} \ ,
\end{equation}
for  clumps with mass  $ M_{clump} \ga 10^{7} M_{\odot}$,
the
lowest resolvable  mass, approximately.
Clumps are more numerous the less massive
they are, thus one would expect a large number of relatively light
clumps. Simulations find  $\alpha \simeq 1.9$ (e.g.,~\cite{Moore1}). We
adopt the same value here and extrapolate to lower masses than those
presently resolved by simulations. With this extrapolation we can
estimate the numbers of clumps in specific mass bins, as
well as the  distances to the nearest of these clumps.

In order to find the proportionality constant in Eq.(\ref{distribution}),
we normalize the mass function so that 500 satellites
exist within the galactic halo with masses above $10^{8} M_{\odot}$, in
agreement with the predictions of N-body simulations for a halo like
ours~\cite{Moore1}. Thus we find for the predicted number, $N_{clump}$,
  of clumps
with mass $M$ greater than $M_{min}$, in solar masses,
\begin{equation}
N_{clump}(M \geq M_{min})= 500 \times \frac{\left(M_{max}/
    M_{min} \right)^{0.9} -1}{\left(M_{max}/ 10^{8} M_{\odot}\right)^{0.9} -1}
\end{equation}
where $M_{max}$ is the maximum mass clump in the halo.
Using our halo as the host within which the clumps
are orbiting,  we assume $M_{max} \simeq 1\% M_{host}
\simeq 2\times 10^{10} M_{\odot}$.

The number, $N_{M_{clump}}$, of clumps predicted in several mass
bins that lie in the mass range  from
$10^{2} M_{\odot}$ up to $10^{8} M_{\odot}$ are
tabulated in the first and second columns of Table \ref{numbers}.
We choose each bin
labeled by $M_{clump}$  to contain clumps with mass
in the range  $[0.5 M_{clump},
5  M_{clump}]$, and calculate
$N_{M_{clump}} = N_{clump}(M \geq 0.5 M_{clump})- N_{clump}(M \geq 5
M_{clump})$.  The numbers appear to be
high enough, and even though they are consistent
with simulations, the question whether they can be
realistic is raised. Contrary to what was initially believed, namely
that large numbers of substructures within a galactic halo  might
threaten the stability of the disk, nowadays we know that this
is not the case (for a recent review, see~\cite{iro02}).
It has been found, first, that the orbits of
satellite substructures in present-day halos very rarely take them
near the disk~\cite{font},
and second, that disk overheating and stability would
become real problems only  when interactions with numerous, large
satellites (e.g., the LMC) take place~\cite{waketal,veletal}.
Consequently, these numbers
can be considered perfectly realistic.
In the third column of Table \ref{numbers} we give values of the mean
inter-clump distance in a specific
mass bin,  assuming that the clumps are homogeneously
distributed within the virial extend of the halo.
Thus, in the third column we have the
quantity $d_{mean}=R_{virial}/ N_{M_{clump}}^{1/3}$, with $R_{virial}$ the
virial radius of the galactic halo which we take to be
equal to 300 kpc. However,
    clumps are unlikely to be homogeneously distributed
in the host halo. It is believed that they are
distributed in a way similar to
the way the smooth component  of the dark
matter halo is distributed (e.g.,~\cite{Pasquale1}).
Thus,  we estimate another measure for the distance,
the distance  to the nearest
clump  for each mass bin, as follows: first,
we calculate the fraction, $f$, of the total dark matter of the galaxy which
is in the form of clumps of mass $M_{clump}$ and take
$f \simeq N_{M_{clump}} M_{clump} / M_{host}$. For
a local dark matter density $\simeq 0.01 M_{\odot} /$pc$^{3}$
and  using
the above fractions we  find the local dark matter density
which is in clumps of a specific mass.
Going from the mass to the number
density, we  find the distance to the nearest clump, $d_{nearest}
\simeq 36 \rm{kpc} / N_{M_{clump}}^{1/3}$ (see also~\cite{Lake1,
Bergstrom2}). As we can see from Table \ref{numbers},
$d_{nearest}$ can be a factor of ten
smaller than the mean distance as calculated
in the homogeneous distribution case.  We use
$d_{nearest}$ below when estimating $\gamma$-ray fluxes from different
clumps.

\begin{table}
\caption{\label{numbers}
Numbers of clumps, mean and nearest clump distances.}
\begin{tabular}{cccc} \hline \hline
$M_{clump}$ & $N_{M_{clump}}$ & $d_{mean}$
& $d_{nearest}$ \\
$(M_{\odot})$ &  & (pc) & (pc) \\ \hline
$10^{2}$ & $2.1 \times 10^{8}$ &  504 & 61 \\
$10^{3}$ & $2.6 \times 10^{7}$ &  1014 & 123 \\
$10^{4}$ & $3.2 \times 10^{6}$ &  2037 & 246 \\
$10^{5}$ & $4.1 \times 10^{5}$ &  4050 & 490 \\
$10^{6}$ & $5.2 \times 10^{4}$ &  8040 & 973 \\
$10^{7}$ & $6.6 \times 10^{3}$ &  16023 & 1939 \\
$10^{8}$ & $8.2 \times 10^{2}$ &  32013 & 3874 \\ \hline \hline
\end{tabular}
\end{table}

One should keep in mind that since we only have one halo to
probe, our galactic halo, the actual distances to the nearest clumps
will have large fluctuations. For a number of realizations of the
clumpy halo in $\gamma$-rays, see \cite{abo02}. Instead of simulating the
sky, we focus on the strategy for future ACT studies based on
$d_{nearest}$. As will be shown,
the results we derive can be easily extended  to
accommodate different clump distances from the ones
we use. A deeper understanding of the clump  space
distribution awaits higher resolution simulations.

\subsection{The structure of the dark matter clumps}
We model the clumps using three different mass density profiles,
$\rho(r)$, the Singular Isothermal Sphere (SIS),
the Moore et al.~\cite{Moore1}, and the Navarro, Frenk
and White (NFW)~\cite{navarro1}
density profile. The SIS profile is described by
\begin{equation}
\rho(r)=\frac{\rho_{o}}{r^{2}},
\end{equation}
and represents a reasonable upper limit to the degree of central
concentration of the clumps. At the center, the SIS
density profile diverges
as $r^{-2}$. The Moore et al. density profile is given
by
\begin{equation}
\rho(r)=\frac{\rho_{o}}{\left({r / r_{s}}\right)^{1.5}
\left[1+ \left({r/ r_{s}}\right)^{1.5} \right]}
\end{equation}
where $\rho_o$ and $r_{s}$ are the
characteristic density and scale radius of the configuration,
respectively.
This profile is the one suggested by numerous N-body
simulations and is used in our
study as representative of the intermediate concentration cases, at least
compared to the  other two density profiles.
It is divergent
at the center where it behaves as $r^{-1.5}$,
whereas at large distances ($r>r_{s}$) it behaves
as $r^{-3}$.
The NFW profile~\cite{navarro1} can be written as
\begin{equation}
\rho(r)=\frac{\rho_{o}}{\left({r / r_{s}}\right) \left({1 + {r
/ r_{s}}}\right)^{2}} \ .
\end{equation}
This profile was the first proposed density
profile for the  dark matter halos produced in simulations.
It is divergent at the center with an
$r^{-1}$ behavior, namely it is  shallower, less centrally concentrated
than both the SIS and the Moore et
al. profiles. At larger radii, NFW  behaves as
$r^{-3}$,  exactly the same way as the Moore et al. profile.
Note that the two quantities, $\rho_{o}$  and $r_{s}$,
in the NFW and Moore et al. profiles are not
the same for the same clump (for a comparison, see~\cite{klypin}).
We choose to use the NFW profile as the least centrally
concentrated density profile. The density profile derived from Taylor and
Navarro (TN)\cite{taylor,taynav},  which is even shallower at the 
center behaving as
$r^{-0.75}$, is a possibility; it will give lower fluxes than the cuspier NFW,
with the latter, as will be seen in what follows, being difficult to 
detect. The
TN will be even more difficult to detect, implying for its 
detectability  neutralino
annihilation cross sections higher than the relevant range of values 
as predicted by
supersymmetric models.

Our final three  choices represent well a range of possible concentrations
of clumps in the galactic halo. In reality, the SIS configuration is 
very hard to
achieve dynamically and has been recently ruled out for most of the clumps in
our halo by EGRET bounds \cite{abo02}. We consider the SIS case to illustrate
the range of clumps more highly concentrated than Moore et al. In 
addition, individual
clumps may vary in concentration, given a range of possible initial 
conditions for a dark
matter overdensity. Thus, an individual clump may be more highly concentrated
than the mean. A possible range in concentrations should be kept in mind in
general searches such as the one we study here.

For a specific clump mass and distance, we need to determine the
parameters of each of the three profiles.
In the case of the SIS, we use the following two
conditions  to calculate $\rho_{o}$ and the radius
of the clump, $R_{clump}$:
\begin{enumerate}
\item The volume integral of the density must yield the mass of the clump:
\begin{equation}
\label{mass}
\int_{0}^{R_{clump}} \rho(r) \ dV= M_{clump}.
\end{equation}
\item The density of the clump at distance $R_{clump}$ from its
center must equal the local density, $\rho_{G}$, of the galactic (host)
    halo at distance
$r_{clump}$,
with $r_{clump}$ the distance of the clump from the
center of the  host halo:
\begin{equation}
\label{joint}
\rho_{G}(r_{clump})=\rho_{clump}(R_{clump}).
\end{equation}
\end{enumerate}

For the case of either the Moore et al. or the NFW density profile,
we
use Eqs.(\ref{mass}) and (\ref{joint}),
but we need a third equation since we have three
unknown quantities  $\rho_o$, $r_{s}$, and $R_{clump}$.
The third condition comes from the requirement that
the radius of the clump be (smaller or) equal to
the tidal radius (see, e.g.,~\cite{binney}).
This guarantees that the clump is not tidally stripped.
Thus,
\begin{equation}
\label{tidal}
R_{clump} = R_{tidal} \simeq r_{clump}
\left[\frac{M_{clump}}{3M_{G}(r_{clump})} \right]^{\frac{1}{3}}
\end{equation}
with $M_{G}(r_{clump})$ the galactic halo mass inside a sphere of radius
$r_{clump}$.
The SIS configuration is also stable against tidal stripping,
as can be seen by considering the similar meaning of
Eqs.(\ref{joint})
and (\ref{tidal}).

In both Eqs.(\ref{joint}) and (\ref{tidal}) we need to specify the
density profile of the galactic halo. Both the Moore et al.
and the NFW profile are acceptable profiles for its description.
Furthermore, for our purposes, they
are essentially  equivalent descriptions.
Convergence studies have shown that these two density profiles
are very similar at radii above $\sim 1 \%$ of the virial radius
(e.g.,~\cite{klypin}). For a virial radius
for our halo $\sim$ 300 kpc, this means that
the two density profiles are different only within the inner
    3 kpc. Given that we are at 8.5 kpc from the galactic center,
and given the nearest clump
distances presented in Table \ref{numbers},
the relevant distances  from the center of the
halo we are concerned with are far larger than 3kpc;
thus, it is not of crucial importance whether the
NFW or the Moore et al.  density profile is
used for the galactic halo~\cite{one}.
We choose the NFW profile to model the
galactic halo, regardless of the density profile assumed for the clumps.
For $r_{s}$ we use 27.7 kpc to match simulation results~\cite{cm01}.
To find $\rho_{oG}$, the characteristic density for
our galaxy, we normalize the density profile so that the peak
circular velocity $v$ is  about 220 km/sec, and thus
\begin{equation}
\rho_{oG}
=\frac{R_{o}v^{2}}{4 \pi r_{s}^{3} G} \left[ ln(1+x_{o}) -
\frac{x_{o}}{1+x_{o}} \right]^{-1}
\end{equation}
with $x_{o}= R_{o} / r_{s}$, and  $R_{o}$ = 8.5 kpc our
galactocentric distance.
Finally,
\begin{equation}
\label{G}
\rho_{G}(r) \simeq \frac{6.18 \times 10^{8}
M_{\odot} /{\rm kpc}^{3}}{\left( {r / {27.7 {\rm kpc}}}\right)
\left[{1 + ({r / 27.7{\rm kpc}})}\right]^{2}}.
\end{equation}

One final characteristic of the dark matter clump structure is the
core radius. Inside some radius $R_{core}$,
the  annihilation rate
becomes so large that the overdensity is destroyed as fast
as it can fill the region, yielding thus a constant density core.
Following~\cite{bgz92}, we find $R_{core}$   by equating the 
annihilation time scale
to the free-fall time scale,
\begin{equation}
\label{core}
\tau_{ann}=\tau_{ff} \rightarrow \left[\frac{\rho(R_{core})}{m_{\chi}}
  \langle  \sigma v \rangle \right]^{-1}=
\frac{\pi}{4} \sqrt{\frac{2R_{clump}^{3}}{GM_{clump}}}.
\end{equation}
In this way we derive a maximum possible $R_{core}$. The actual value
might be smaller. Note that the choice of $R_{core}$ affects only the 
results for the SIS density
profile since, as can be easily verified for the simple case of an 
unresolved clump,
the  flux coming from such a clump scales as $R_{core}^{-1}$ (see section
\ref{fluxsection}).  In the case of
the Moore et al. density profile the $R_{core}$-dependence of the flux
is via the factor $0.67(1/x_{clump}^{1.5} -1) -0.005\log{x_{clump}}-\log{x_{core}}$,
approximately, with $x_{clump}=R_{clump}/r_{s}$ and 
$x_{core}=R_{core}/r_{s}$. Namely, the
Moore et al. conclusions are not sensitive to the exact choice of 
$R_{core}$~\cite{comment}.
In the case of the core-dominated fluxes of an SIS clump, using the 
maximum $R_{core}$
will yield the minimum fluxes. As will become clear from our results 
however, the
conclusions do not change, at least qualitatively, since the SIS is 
steep enough to be
detectable even when using these minimum fluxes.

Below we discuss the annihilation cross sections into
either  continuum or monochromatic
$\gamma$-rays. We find that
the dominant cross section times relative velocity product
corresponds to annihilation into the continuum, $\langle \sigma v
\rangle_{cont.}$, and even though, in principle, all
annihilation channels contribute in smearing out the central cusp,
we use this dominant cross
section  in Eq.(\ref{core}) to find $R_{core}$.

\section{Neutralino annihilation as a gamma-ray source}

\subsection{The Neutralino-Supersymmetric models for dark matter}

Following~\cite{Bergstrom2,Baltz1,Bergstrom3,Edsjo1},
we work in the frame of the Minimal Supersymmetric extension
of the Standard Model (MSSM) using the
computer code DarkSUSY~\cite{darksusy1}.
The general R-parity conserving version
of this model is characterized by more than a hundred free parameters.
It is of common praxis to
make some simplifying  assumptions which leave only 7
parameters, the higgsino mass parameter $\mu$, the gaugino mass parameter
$M_{2}$, the ratio of the Higgs vacuum expectation values given by $\tan
\beta$, the mass of the CP-odd Higgs boson $m_{A}$, the scalar mass
parameter
$m_{o}$, and the trilinear soft SUSY-breaking parameters
$A_{t}$ and $A_{b}$ for third generation quarks
(for a discussion of the model and
the simplification  procedure, the results of which DarkSUSY
incorporates, see~\cite{Bergstrom4,Bergstrom8,Bergstrom9}).

The LSP  in most
models is the lightest of the neutralinos. Neutralinos
are a linear superposition  of four neutral spin-$1/2$
Majorana particles: the superpartners of the Higgs, namely
the neutral, CP-even
higgsinos ${\tilde {H_{1}}}^{0}$ and ${\tilde{H_{2}}}^{0}$, and the
superpartners of the electroweak
gauge bosons, namely the bino $\tilde{B}$ and the
wino ${\tilde{W}}^{3}$. After
electroweak symmetry breaking, these gauge eigenstates mix. The
diagonalization of the corresponding tree-level mass matrix  gives the
neutralinos,
\begin{equation}
\nonumber
\tilde{\chi_{i}}^{0}=N_{i1} \tilde{B} +N_{i2} {\tilde{W}}^{3} +
N_{i3} {\tilde{H_{1}}}^{0} +N_{i4} {\tilde{H_{2}}}^{0}.
\end{equation}
The lightest of these, the  $\chi={\tilde{\chi_{1}}}^{0}$,  is assumed
to be the LSP and is referred to as {\it the} neutralino.
Due to R-parity conservation the neutralino is stable, since there is
no allowed state for it to decay into. Its mass is somewhere between
some tens of GeV up to several TeV. It is cold, namely non-relativistic
when decoupled, and is considered one of the most plausible
candidates for the CDM particle.

Using DarkSUSY, we made random scans of the  parameter space,
with overall limits for the seven MSSM parameters as given in
Table \ref{susy}.
For the neutralino and chargino masses, one-loop corrections
were used in accordance to~\cite{hfast1},
whereas for the Higgs bosons the leading log two-loop
radiative corrections  were calculated using the code
FeynHiggsFast~\cite{hfast} incorporated in DarkSUSY.
\begin{table}
\caption{\label{susy} The ranges of parameter values used in the scans of
the SUSY parameter space.}
\begin{tabular}{cccccccc}  \hline \hline
Parameter & $\mu$ & $M_{2}$ & $\tan{\beta}$ & $m_{A}$ & $m_{0}$ &
$A_{b}/m_{0}$
& $A_{t} / m_{0}$ \\
unit & GeV & GeV & 1 & GeV & GeV & 1 & 1  \\ \hline
Min & $-50000$ & $-50000$ & 3.0 & 0 & 100 & $-3$ & $-3$ \\
Max & $+50000$ & $+50000$ & 60.0 & 10000 & 30000 & $+3$ & $+3$ \\ \hline
\hline
\end{tabular}
\end{table}

We scanned the parameter space to obtain the boundaries of the allowed
SUSY parameter space which we plot in FIGs.1-3, 5, and 7. For each
model, we checked whether it is excluded by accelerator
constraints. We used the latest
constraints available in the DarkSUSY package, namely the 2000
constraints which are adequate for our purposes. The most
important constraints come from the LEP collider at CERN,
with respect to the lightest chargino and the lightest Higgs boson mass,
as well as constraints from
$b \rightarrow s\gamma$ (for more details see, e.g.,~\cite{Baltz1}
and references therein).
For each model consistent with the accelerator constraints, we
calculated the corresponding  $\Omega_{\chi} h^{2}$, with
$\Omega_{\chi}$ the neutralino relic density in units of the critical
density, and $h$  the Hubble parameter  today  in units of 100
km s$^{-1}$ Mpc$^{-1}$. DarkSUSY calculates the relic density solving the
Boltzmann equation numerically, while taking into account resonances and
thresholds, and all the tree-level 2-body annihilation channels
(see,~\cite{Edsjo1, Gondolo1}).
We take into account the coannihilations between
the lightest neutralino  and the heavier neutralinos/charginos
only if the mass difference is less than $30 \%$,
obtaining thus  a relic density with accuracy $\simeq 5\%$,
which is reasonable for our purposes.
Lastly we are interested in models in which the
corresponding relic abundance  contributes a significant, but not excessive
amount to the overall  density. We choose the range
\begin{equation}
\label{omega}
0.1 \leq \Omega_{\chi}h^{2} \leq 0.3
\end{equation}
to be consistent with cosmological constraints.
The lower limit in inequality (\ref{omega})  is chosen on the
basis that for values lower than 0.1 there is not enough dark matter
to play a significant role in structure formation, or to constitute
a large fraction of the critical density.  The upper bound
is chosen on the basis that, given that there is a minimum  age
for the Universe, there is a maximum value that the dark matter density
can have. The upper limit 0.3 corresponds to a minimum age of 12 Gyr
approximately (see, e.g.,~\cite{Peacock1}).
Note that the choice of
$0.3$  for the upper limit can be considered generous, especially
given the results from recent SN observations~\cite{reiss1,Perlmutter1},
which constrain the allowed relic density to about
$\Omega_{\chi} h^{2} \simeq 0.15$, with $h^{2} \simeq 1/2$.

\subsection{Gamma-rays from neutralino annihilation}

Neutralinos annihilate through a variety of channels. We
will focus  on the  continuum $\gamma$-rays, and on the
two monochromatic lines that are produced
through the cascade decays of other
primary annihilation products.
The continuum contribution is mainly due to  the decay of
$\pi^{0}$ mesons produced in jets from neutralino annihilations.
Schematically,
\begin{eqnarray}
\label{anni}
\chi+\bar{\chi} & \rightarrow & \pi^{0}    \nonumber \\
                  &             & \hookrightarrow   \gamma+\gamma.
\end{eqnarray}
To model the fragmentation process and extract information on the
number and energy spectrum of the $\gamma$-rays produced, we adopt
a simplified version of the Hill spectrum~\cite{hill1,hill2} for
the total hadron spectrum  produced by quark
  fragmentation based on the leading-log approximation (LLA),
\begin{equation}
\frac{dN_{h}}{dx_h} \simeq \frac{15}{16} x_h^{-3/2} (1-x_h)^{2}
\end{equation}
where $x_h=E_{h}/m_{\chi}$, and $E_{h}$ is the energy of
a hadron in a jet with total energy equal to the neutralino
mass, $m_{\chi}$. Assuming that all the hadrons produced are pions, which
is very close to reality, and assuming that each pion family takes
approximately $1/3$ of the total pion content of each jet we may write
for the
$\pi^{0}$ spectrum,
\begin{equation}
\label{spec1}
\frac{dN_{\pi^{0}}}{dx_{\pi}} \simeq \frac{5}{16} x_{\pi}^{-3/2}
(1-x_{\pi})^{2}
\end{equation}
with $x_{\pi}=E_{\pi^{0}} / m_{\chi}$.
Furthermore, the probability per unit energy that a neutral pion with energy
$E_{\pi^{0}}$  produces a photon  with energy $E_{\gamma}$
through the process shown in Eq.(\ref{anni}) is $2/E_{\pi^{0}}$.
Thus, from Eq.(\ref{spec1}) we get for the continuum
photon spectrum,
\begin{equation}
\label{gspectrum}
\frac{dN_{cont.}}{dx_{\gamma}} \simeq \int_{x_{\gamma}}^{1} \frac{2}{y}
\frac{dN_{\pi^{0}}}{dy} dy
\end{equation}
with $x_{\gamma}={E_{\gamma}/m_{\chi}}$ and $y={E_{\pi^{0}}/m_{\chi}}$.

The continuum signal lacks distinctive features; this might make it
difficult to discriminate from other possible $\gamma$-ray sources. A
more unique signature is given by monochromatic $\gamma$-rays, which
arise from loop-induced s-wave annihilations into the $\gamma\gamma$
and the
$Z\gamma$ final states. These lines are free of  astrophysical
backgrounds. The respective {\it spectra} for
the
two lines are:
\begin{itemize}
\item For the $\gamma\gamma$, 2 photons at $E_{\gamma}\simeq
m_{\chi}$
\item For the $Z\gamma$, 1 photon at
$E_{\gamma} \simeq m_{\chi}- {m_{Z}^{2} / {4 m_{\chi}}}$,
\end{itemize}
    as  can be easily verified by the conservation laws and the fact that
the neutralinos are expected to be highly non-relativistic.

Note that for high $m_{\chi}$ values, the energies of the two lines are
expected to coincide. Furthermore, in the case of the $Z\gamma$ line,
the energy of the photon becomes vanishingly small as
$m_{\chi}$ tends to $m_{Z}/2$. Below $m_{\chi}={m_{Z}/2}$, the
$Z\gamma$ process is no longer kinematically allowed.

\section{The gamma-ray signal}
\label{fluxsection}
\subsection{Fluxes and counts}

We derive the fluxes for resolved and unresolved clumps as a function of
the neutralino parameters, and the size and distance of the clumps, assuming
the nearest clump for each mass bin.
To
turn the estimated fluxes into photon counts collected
by an ACT, we need to specify
the characteristics of the particular ACT. We choose to use VERITAS as
the standard next generation ACT and list
its specifications in Table
\ref{veritaspecs}.

\begin{table}
\caption{\label{veritaspecs} VERITAS specifications}
\begin{tabular}{lcl} \hline  \hline
Energy range  & 50 GeV-50 TeV & \\
Energy resolution & $15 \%$ & \\ \hline
Effective area &   Angular resolution & \\  \hline
$1 \times 10^{8}$ cm$^{2}$  & $5^{'}$  & at 100 GeV \\
$4 \times 10^{8}$ cm$^{2}$ & $3^{'}$  & at 300 GeV  \\
$1 \times 10^{9}$ cm$^{2}$ & $2^{'}$  & at 1 TeV \\ \hline \hline
\end{tabular}
\end{table}

The number, $R$, of annihilations per unit volume and time is
\begin{equation}
R= n^{2} \langle\sigma v\rangle_{i}
\end{equation}
where $n$ is the number density of neutralinos, and $\langle \sigma
v\rangle_{i}$ is the thermally averaged product of
the cross section times
the relative
velocity~\cite{two}
for the continuum $\gamma$-rays ($i$=cont.), the 2$\gamma$ line ($i$=$\gamma
\gamma$), and the $Z\gamma$ line ($i=Z\gamma$).
Denoting by  $N_{i}$ the number of $\gamma$-rays emitted per
annihilation, with energy $E_{\gamma}$
above the energy threshold $E_{th}$ used, then
$N_{\gamma\gamma}=2$ (provided that $m_{\chi} \ge  E_{th}$), and
$N_{Z\gamma}=1$
(provided that $m_{\chi}- {m_{Z}^{2} / {4 m_{\chi}}} \ge E_{th}$);
$N_{cont.}$
can be obtained by integrating the spectrum given
by Eq.(\ref{gspectrum}),
\begin{equation}
N_{cont.}(E_{\gamma} \ge E_{th})=
  \int_{x_{th}}^{1} \frac{dN_{cont.}}{dx_{\gamma}}dx_{\gamma}
=\frac{5}{6}x_{th}^{3/2}-\frac{10}{3}x_{th} + 5\sqrt{x_{th}}
+\frac{5}{6 \sqrt{x_{th}}} -\frac{10}{3} \ .
\end{equation}
where  $x_{th}={E_{th}/m_{\chi}}$.

The emission coefficient $j_{i}$, namely the
number of emitted photons per unit
time and volume, will be
\begin{equation}
j_{i}= N_{i} n^{2} \langle\sigma v\rangle_{i} =
\frac{N_{i} \rho^{2}
\langle \sigma v\rangle_{i}}{m_{\chi}^{2}}
\end{equation}
where $\rho$ is the mass density of neutralinos.

Given the expression for the emission coefficient, we can
calculate the flux coming from a clump. The observed flux
at the Earth depends on whether the object is resolved or
unresolved by the ACT, namely, whether the angular size of the
clump is larger or smaller than the angular
resolution of the instrument. More specifically, in the case
where the clump under consideration is unresolved, its flux at
the Earth, $F_{unres}^{i}$, is given by the volume integral of the emission
coefficient divided by the area $4 \pi d^2$,
\begin{equation}
F_{unres}^{i}=\frac{1}{4 \pi d^{2}} \int_{0}^{R_{clump}}  j_{i} d^{3}R=
\frac{1}{4 \pi d^{2}}
\frac{N_{i} \langle\sigma v\rangle_{i}}{m_{\chi}^{2}}
\int_{0}^{R_{clump}} \rho^{2} d^{3}R
\end{equation}
where $d$ denotes the distance of the clump from the Earth and
$R_{clump}$ is, as before, the clump radius.
At this point,
note that  $d$ and $r_{clump}$ (appearing in Eqs.(\ref{joint})   and
(\ref{tidal})) are two different quantities; $d$ is the distance
of the clump from the Earth, whereas $r_{clump}$ is the distance
of the clump from the center of the Milky Way. For a clump with
galactic coordinates ($l$,$b$), the two quantities are related via
the expression,
\begin{equation}
\label{distance}
r_{clump}^{2}=d^{2} + R_{o}^{2} - 2dR_{o} \cos{l}\cos{b}
\end{equation}
with $R_{o}=$8.5 kpc.
If the clumps
are resolved, the corresponding flux at the Earth, $F_{res}^{i}$,
is given by the integral
    along the line of sight of the emission  coefficient,
\begin{equation}
\label{resolved}
F_{res}^{i}=\frac{1}{4 \pi}
\int_{los} j_{i} dz=\frac{1}{4 \pi} \frac{N_{i} \langle \sigma
v\rangle_{i}}{m_{\chi}^{2}}
\int_{los} \rho^{2} dz
\end{equation}

In the case of the unresolved sources, the source counts,
$N_{unres}^{i}$, will be
simply,
\begin{equation}
N_{unres}^{i}= F_{unres}^{i} A_{eff} t
\label{unres}
\end{equation}
with $A_{eff}$ the effective collecting area of the instrument
and $t$ the integration time.
For a resolved clump, we can define its surface brightness
$\mu^{i}$ as seen by an ACT as
\begin{equation}
\label{brightness}
\mu^{i}= F_{res}^{i} A_{eff} t
\end{equation}
where $A_{eff}$ and $t$ the same as above.
To obtain the number of photons,  $N_{res}^{i}$, that
the instrument will collect from the source,
assuming that we observe the central pixel, we integrate the surface
brightness of the source over a  disk
centered on the source  with angular radius equal to the
angular resolution of the instrument $\sigma_{\theta}$,
\begin{equation}
\label{expre}
N_{res}^{i}
=\int_{0}^{2 \pi} d\phi \int_{0}^{\sigma_{\theta}}
\mu^{i} \theta   d\theta.
\end{equation}
The angular resolution of the instrument defines a maximum
projected size $R_{max}$ for a clump at distance $d$ by
$\sigma_{\theta}={R_{max} / d}$. Similarly, $\theta= {R/ d}$.
Substituting Eqs.(\ref{resolved}) and
(\ref{brightness}) in Eq.(\ref{expre}) we  obtain,
\begin{equation}
\label{source}
N_{res}^{i}=\frac{1}{4 \pi d^{2}}
\frac{N_{i} \langle \sigma v\rangle_{i}}{m_{\chi}^{2}} A_{eff}t
\int_{0}^{2 \pi} d\phi \int_{0}^{R_{max}}R \, dR \int_{los}
\rho^{2} dz.
\end{equation}

We find that most clumps we considered
will be resolved by future ACTs, e.g., VERITAS~\cite{unre}.
Thus,
we focus on the resolved fluxes and counts when discussing our results in \S V.
  For the diffuse $\gamma$-ray  contribution of clumps
in the halo of our galaxy
see \cite{cm01,abo02}.

\subsection{Background counts}
The detectability of the clumps depends on their fluxes as compared to
the possible background contributions. As we discuss below, the dominant
background contributions are
due to electronic and hadronic cosmic ray showers.
Efforts to discriminate between photon primaries versus charged primaries
can help improve considerably the detectability of dark matter
   clumps. We also
discuss the galactic and extragalactic $\gamma$-ray backgrounds, as well as
the contribution from the rest of the dark matter halo component.

\subsubsection{Electronic and hadronic cosmic ray showers.}

For the background rate of $\gamma$-like hadronic
showers we use the integral spectrum  presented
in~\cite{Bergstrom6},
\begin{equation}
\label{hadro}
\frac{dN_{h}}{d \Omega} (E >E_{th}) =
6.1 \times 10^{-3} \left(\frac{E_{th}}{1 {\rm GeV}} \right)
^{-1.7} {\rm cm}^{-2}{\rm s}^{-1}{\rm sr}^{-1}.
\end{equation}
This was derived from data taken with the Whipple 10m telescope.
VERITAS is expected to have an improved method
for rejecting charged primary showers,  compared to the
Whipple telescope.  However, aiming at a conservative treatment, we do not
take this into account.
For the electron-induced background we use the following
integrated spectrum~\cite{Longair},
\begin{equation}
\label{ele}
\frac{dN_{e^{-}}}{d \Omega}(E >E_{th}) =
3.0 \times 10^{-2} \left(\frac{E_{th}}{1{\rm GeV}} \right)^{-2.3}
{\rm cm}^{-2}{\rm s}^{-1}{\rm sr}^{-1}.
\end{equation}
The electron-induced showers at  Whipple are indistinguishable from
$\gamma$-rays and can be rejected only on the basis of their
arrival direction, whereas the hadron-induced showers are
more extended on the ground than the electron-induced ones.
Furthermore, we see that cosmic-ray electrons have a steeper spectrum than
cosmic-ray nuclei and dominate the background
at low energies.

\subsubsection{Extragalactic and galactic $\gamma$-ray emission.}

For the extragalactic diffuse emission, we use a fit to the EGRET data,
valid  for
the energy range from 30 MeV to $\sim$ 100  GeV \cite{Sreekumar}
\begin{equation}
\label{extra}
\frac{dN_{eg}}{d\Omega dE}=
(7.32 \pm 0.34)\times 10^{-9} \left(\frac{E}{451 {\rm MeV}}\right)
^{-2.10\pm 0.03} {\rm MeV}^{-1}{\rm cm}^{-2}{\rm s}^{-1}{\rm sr}^{-1}
\ .
\end{equation}

The galactic diffuse emission is thought to be mainly due to
cosmic-ray protons and electrons interacting with the interstellar medium.
It  is enhanced towards the galactic center and the galactic disk, and
has been measured by EGRET~\cite{Sreekumar,Hunter}
up to about 10 GeV. For its differential spectrum we use the
expression provided in~\cite{Bergstrom6}, namely
a power law fall-off in energy of the form
\begin{equation}
\frac{dN_G}{dE}(E,l,b)=  10^{-6} N_{0}(l,b)
\left(\frac{E}{1 {\rm GeV}}\right)^{\alpha}
{\rm GeV}^{-1} {\rm cm}^{-2}{\rm s}^{-1} {\rm sr}^{-1}.
\end{equation}
The factor $N_{0}$ depends only on the galactic coordinates
$(l,b)$ and has been fixed using EGRET data at 1 GeV (see~\cite{Bergstrom6},
and references therein). It is given by
\begin{eqnarray}
N_{0}(l,b)& = &{\frac{85.5}{\sqrt{1+(l/35)^{2}}
\sqrt{1+(b/(1.1 +|l|0.022))^{2}}}}
+0.5  \hspace{0.2cm} if \hspace{0.1cm} |l| \geq 30^{\circ} \nonumber  \\
& = & {\frac{85.5}{\sqrt{1+(l/35)^{2}}
\sqrt{1+(b/1.8)^{2}}}}
+0.5  \hspace{0.2cm} if \hspace{0.1cm} |l| < 30^{\circ}
\end{eqnarray}
with $l$ and $b$  varying in $[-180^{\circ},180^{\circ}]$ and
$[-90^{\circ},90^{\circ}]$, respectively.
This expression is in reasonable agreement with data especially
towards the galactic center. The exponent $\alpha$   in
principle depends on both the energy range and the galactic coordinates; as
in ~\cite{Bergstrom6,Bergstrom7}, we use $\alpha=-2.7$, assuming the
behavior continues at higher energies.

It is also important to obtain an estimate of the  $\gamma$-ray
counts due to neutralino annihilation in the smooth dark matter component of
the galactic halo. The flux of continuum $\gamma$-rays
of the smooth halo  component is given by
\begin{equation}
F_{smooth}(\psi)=
\frac{1}{4 \pi} \frac{N_{cont.} \langle \sigma v\rangle}{m_{\chi}^{2}}
\int_{los} \rho_{G}^{2}(l) dl(\psi).
\end{equation}
with $\rho_{G}$  given by Eq.(\ref{G}), and
$\psi$ the  angle of the direction of observation
with respect to the direction of the
galactic center.  This flux is clearly enhanced towards
the central regions of the galaxy  (for more details see,
e.g.,~\cite{cm01}). In order to estimate a value close to
the maximum possible smooth component counts, $N_{smooth}$,
    we use the direction  $\psi=10^{o}$ -- close to the center.
The  flux coming from this direction is approximately
\begin{equation}
F_{smooth} (E>E_{th}) \simeq
3.5 \times 10^{-26} N_{cont.} \frac{\langle\sigma v\rangle_{cont.}/ \rm{cm}^{3}\rm{s}^{-1}}{m_{\chi}^{2}/\rm{g}^{2}} 
\  {\rm s}^{-1}{\rm cm}^{-2}{\rm sr}^{-1}.
\end{equation}

The emission due to  the total clumped component of the dark matter halo,
as seen in \cite{cm01,abo02},
overwhelms the smooth component $\gamma$-ray emission.
Note that neither the smooth, nor the clumped component
$\gamma$-ray emission constitutes an additional background for our study.
For example, part or all of the extragalactic background, which
we have already taken into account, would be due to numerous,
unresolved  clumps in our halo.
In fact, this has allowed ref.~\cite{abo02} to place strong
limits on the neutralino parameter space for
clumped halos with SIS and Moore et al. clump profiles,
by comparing the clumped component $\gamma$-ray emission with
the extragalactic diffuse $\gamma$-ray emission
(Eq.(\ref{extra})). In particular, an SIS
clump distribution would surpass EGRET bounds for
all of the SUSY models usually
considered for neutralinos.

In order to estimate the relative importance of  the
several contributions we assume an
ACT observation with the following characteristics:
$A_{eff}=10^{8}$cm$^{2}$, $t=100$ hrs, $E_{th}=50$ GeV,
$\sigma_{\theta}=0.1^{\circ}$.  For these parameters, we find the counts
$N_{e-}\simeq 1278$, $N_{h} \simeq 2718$, $N_{eg}
\simeq 6$, $N_{G}\simeq 225$ (for $m_{\chi}=100 $ GeV and,
referring to the continuum
$\gamma$-rays, for a typical cross section $\langle \sigma
v\rangle_{cont.} =5 \times 10^{-27} {\rm cm}^{3} {\rm s}^{-1}$ we obtain
$N_{smooth} \simeq 0$, rounding off to the
closest integer). To calculate the galactic background counts, we made the
conservative assumption of the maximum possible background and, thus, used
$(l,b)=(0^{\circ},0^{\circ})$.
This background is anisotropic and will be lower at
higher galactic latitudes.
Evidently, the main backgrounds for ACT searches
are the hadronic and electronic backgrounds. Only if the
composition of primaries can be identified, the galactic and the
extragalactic diffuse $\gamma$-ray backgrounds become important;
this would be highly desirable, given that the galactic background is
(at least) an order of magnitude lower than
the cosmic ray induced backgrounds.

We chose to keep a conservative estimate of the backgrounds
  by considering the
two dominant backgrounds, the electronic and the hadronic
cosmic-ray induced ones, since, as we have shown above,
the other contributions are at least one  order of magnitude lower.
Thus, the background counts in the
case of the continuum are
\begin{equation}
\label{conti}
N_{b}(E \geq E_{th})=\frac{dN_{b}}{d \Omega}(E \geq E_{th})  \Delta\Omega \
A_{eff} \ t
\end{equation}
where we define ${d N_{b} / d \Omega}$ as the sum
of the hadronic and the electronic
cosmic ray  induced backgrounds,
as given by Eq.(\ref{hadro}) and (\ref{ele}),
respectively. In the case of the lines the background counts are given by
\begin{equation}
\label{line}
N_{b}(E=E_{o}) =\left[\frac{dN_{b}}{d \Omega}(E \geq E_{1})-
\frac{dN_{b}}{d \Omega}(E \geq E_{2})\right]\ \Delta\Omega
\  A_{eff} \  t
\end{equation}
where $E_{o}$ stands for the energy of the line and $E_{1},E_{2}$
stand for $(1-R_{E})E_{o}$ and $(1+R_{E})E_{o}$ respectively,
with $R_{E}$ denoting the energy resolution of the detector.

Note that the only anisotropic background -- the
galactic -- is negligible in comparison with the charged particle
backgrounds. This makes our calculations essentially independent of the
exact  direction  of the clump
in the sky. Any particular choice of galactic
coordinates would not limit the generality of our conclusions with respect
to the observability of the clumps. The
only assumption we should make is that the clumps
are more likely to be found away from the galactic plane.
This is to be expected since simulations  show that
the orbits of CDM satellites in present-day  halos very rarely take them
near the disk (e.g.,~\cite{font}).
This is not a very constraining assumption -- all directions in the
sky that avoid the
galactic disk are essentially equivalent -- and thus,
the only  really important quantity
with respect to the location of a clump is its distance from us.
The exact galactic coordinates can in
principle be important  via Eqs.(\ref{joint}) and (\ref{distance}), but
it turns out that this is not the case.
For example, considering the clumps with
mass  $M_{clump} = 10^{8} M_{\odot}$ and nearest
distance of about $4$ kpc (see Table \ref{numbers}),
we find
that  putting  such a clump towards the galactic center yields  $\rho_{G}
\simeq 2.8 \times 10^{9} M_{\odot}$ kpc$^{-3}$, whereas putting
it in the exact
opposite direction,  towards the anticenter,
yields $\rho_{G} \simeq 1.3 \times 10^{9} M_{\odot}$ kpc$^{-3}$, namely
there is no significant
variation.

\section{Results}

\subsection{The detectability condition}

The background follows Poisson statistics and as such it exhibits
fluctuations of amplitude $\sqrt{N_{b}}$. For given technical
and observation parameters, the minimum detectable flux of
$\gamma$-rays for an ACT is determined by the condition that
the significance $S$ exceeds a specific number, $M_{s}$,
of standard deviations, $\sigma$,
or that the number of detected photons exceeds a specific
number $N_{o}$.  Namely,
\begin{equation}
\label{criterion}
S=\frac{N_{s}}{\sqrt{N_{b}}} \geq M_{s}
\end{equation}
and,
\begin{equation}
N_{s} \geq N_{o}.
\end{equation}
The first condition simply  defines what achieving an $M_{s}$-$\sigma$
detection requires. The second condition ensures that
we obtain sufficient statistics of $\gamma$-ray events and is
useful especially at the higher energies ( $\geq 1$ TeV),
where the backgrounds are negligible, and thus the first condition
is useless.  The second condition also shows that
at high energies, whether
the source will be detectable or not is determined  only from its
high energy spectrum (and the  ACT characteristics).
Given that we start at tens of
GeV energy thresholds, we check for the first condition making sure that
the second condition is satisfied as well. We require
  a 5-$\sigma$ detection level ($M_s = 5$). As has been already mentioned,
for all the mass bins
studied, the nearest clump can be
resolved for standard
ACT angular resolutions
(see Table \ref{veritaspecs}),
and thus we use Eq.(\ref{source})  for $N_s$,
whereas for the
background we use
either Eq.(\ref{conti}) or Eq.(\ref{line}), depending on whether we
study the continuum or the lines, respectively.

Using the aforementioned  detectability condition, we can constrain  the
SUSY parameter space. In addition to the unknown neutralino parameters,
$\langle \sigma v\rangle_{i}$ and
$m_{\chi}$, we do not know the precise density profiles
of the  dark matter clumps. In the figures that will follow, we
take under consideration
the profile  uncertainties by using  the
three density profiles
mentioned before (the SIS, Moore et al., and NFW profiles). We
then set the
constraints on the neutralino parameter space as follows: we substitute
Eq.(\ref{source})  for
$N_{s}$, and  Eq.(\ref{conti}) or (\ref{line}) for
$N_{b}$,  as well as the expressions for
$N_{i}$ in condition (\ref{criterion}), and  then we solve the resulting
inequality with respect to $\langle\sigma v\rangle_{i}$. This
yields an inequality of the form
\begin{equation}
\label{constraint}
\langle\sigma v\rangle_{i} \geq f(I,A_{eff},t,E_{th},{dN_{b}/d \Omega},
\Delta \Omega, M_{s}) m_{\chi}^{2}.
\end{equation}
In other words,  an $M_{s}$-$\sigma$ detection  for given
$A_{eff}$, $t$, $E_{th}$, ${dN_{b}/d \Omega}$,
$\Delta \Omega$, and $M_{s}$ can be obtained as long as  $\langle\sigma
v\rangle_{i}$ is above a certain value which depends on $m_{\chi}^{2}$ and
the integral of the density squared, which we have denoted by $I$.
Thus, plotting
inequality (\ref{constraint}) with the equality sign onto the SUSY
parameter space, we
divide the SUSY parameter space into the detectable (above the
line defined by (\ref{constraint})) and the undetectable
(below the line defined by (\ref{constraint})) regions.

In another set of figures, we show the flux for a given clump as a
function of $m_{\chi}$,  assuming a typical value,
$\langle\sigma v\rangle_{typ}$,
for
the cross-section.
This is interesting  to be compared
to the dominating backgrounds, as well as to what we call the minimum
detectable flux $F_{min}$, which we define using again
the detectability criterion given in  (\ref{criterion}). More
specifically, for the minimum required source counts
$N_{s,min}$, so that an $M_{s}$-$\sigma$ detection level be achieved we
will have  from (\ref{criterion})
\begin{equation}
    N_{s,min} =M_{s} \sqrt{N_{b}}
\end{equation}
For $N_{s,min}=F_{min}A_{eff}t$ we obtain,
\begin{equation}
\label{minflux}
F_{min}=\frac{M_{s} \sqrt{N_{b}}}{A_{eff}t}
\end{equation}
This is separately applicable for the continuum and the lines.

\subsection{Continuum $\gamma$-rays}

\begin{figure}
\epsfig{file=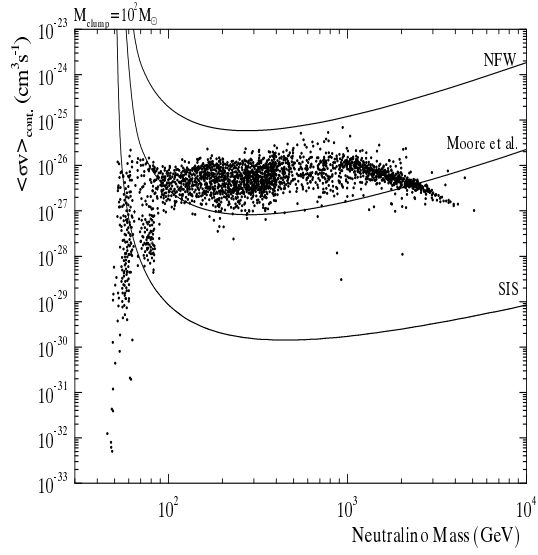, width=8.6cm, height=8.6cm}
\epsfig{file=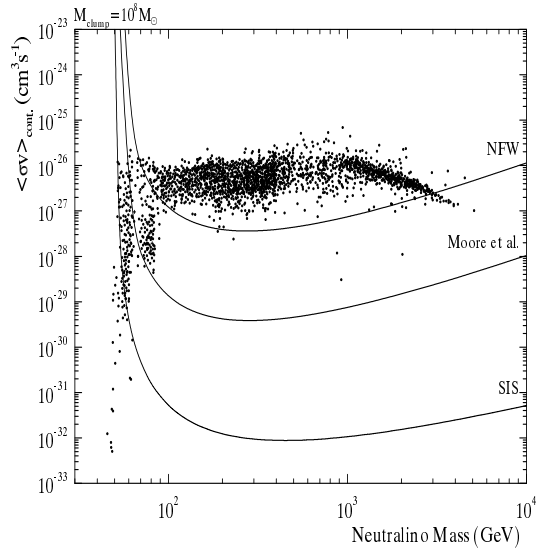, width=8.6cm, height=8.6cm}
\caption{\label{conti-susy}
The minimum detectable $\langle \sigma v \rangle_{cont.}$  versus
$m_{\chi}$ for the SIS, the Moore et al., and the NFW profile.
The clump masses used are $10^{2}
M_{\odot}$ and $10^{8} M_{\odot}$.
The dots represent allowed SUSY models (see
text for details). The lines represent  the
5-$\sigma$  detection for  $A_{eff}=10^{8}$ cm$^{2}$,
$E_{th}$=50 GeV,
$\sigma_{\theta}=6^{'}$, and for 100 hrs of observation.
Only  SUSY models that lie above the corresponding curve will
yield a detectable signal. }
\end{figure}

The minimum detectable neutralino annihilation cross section
into continuum $\gamma$-rays as a function of neutralino mass for
a $10^{2}M_{\odot}$ and a $10^{8}M_{\odot}$
dark matter clump is shown in FIG.\ref{conti-susy}. These
curves were obtained using condition (\ref{constraint})
for $A_{eff}=10^{8}$ cm$^{2}$,
$t=3.5 \times 10^{5}$ s (100 hrs), $E_{th}=50$ GeV and
$\sigma_{\theta}=6^{'}$ (these are also
the parameters we use for the rest of the
figures, unless otherwise stated). In the same figure, the three cases of
clump modeling using an NFW, a Moore et al., or an SIS density profile
are shown. Note that by plotting  the results for the minimum and the
maximum clump masses that we considered,
$10^{2} M_{\odot}$ and $10^{8} M_{\odot}$, respectively,
any other clump mass choice lies
in  between. The same is true for the density profiles, at least
with respect to the degree of  central concentration: any profile steeper
than the NFW and softer than the SIS will be a curve between the two
limiting curves defined for a specific clump mass,
  by the SIS and the NFW profiles,
and as an example we plot the Moore et al. case.

In the case of a $10^{2}M_{\odot}$ clump,
we see that if the clump were described by a density
profile as centrally concentrated as the SIS, then it would
be detectable for most neutralino parameters, unless $m_{\chi} \la$ 70 GeV.
The same clump modeled with the Moore et al. profile
has good chances to be detected, since for a large range of $m_{\chi}$ the
$\langle\sigma v\rangle_{cont.}-m_{\chi}$ curve is below most
of the possible models. Furthermore, note that given the recent
results from \cite{abo02}, these
highly centrally concentrated clumps would have to be a rare event among
concentrations of halo clumps.

The same results are plotted for a clump with mass $10^{8}M_{\odot}$. The
minimum detectable $\langle\sigma v\rangle_{cont.}$ for all the three profiles
goes down by approximately two orders of magnitude, compared to the $10^{2}
M_{\odot}$ case. As a result, such a  massive
clump is  most likely detectable, even if it is best
described by the least concentrated
density profile, the NFW. Note also that, referring to the low mass clumps,
  the NFW case is, compared to the other
density profiles,
the most consistent profile
with EGRET bounds~\cite{abo02}.

\begin{figure}
\epsfig{file=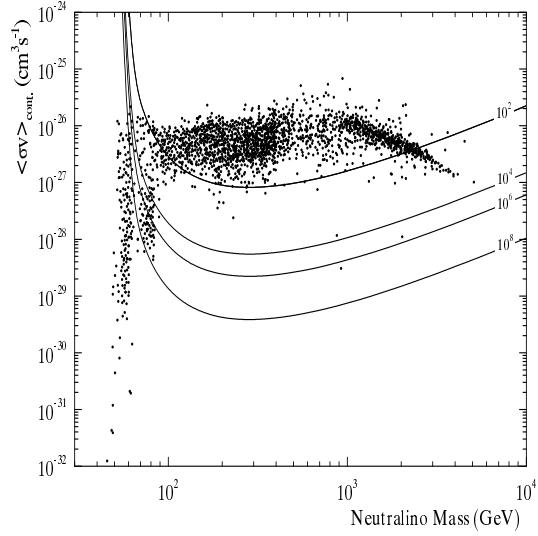, width=8.6cm, height=8.6cm}
\caption{\label{cmass} The minimum detectable $\langle \sigma v
\rangle_{cont.}$  versus
$m_{\chi}$ for Moore et al. clumps
with $M_{clump}= 10^{2} M_{\odot}$, $10^{4} M_{\odot}$, $10^{6} M_{\odot}$
and $10^{8} M_{\odot}$. The dots and the other parameters
are as in FIG.\ref{conti-susy}.
}
\end{figure}

The way the constraints imposed on the  SUSY
parameter space depend on the mass of the clump is
depicted in FIG.\ref{cmass},
assuming a Moore et al. density profile, and
   for the same observation parameters as in FIG.\ref{conti-susy}.
The results are plotted for four different clump masses: $10^{2} M_{\odot}$,
$10^{4} M_{\odot}$, $10^{6} M_{\odot}$ and $10^{8} M_{\odot}$.
As evident, the higher mass clumps have better chances of detectability,
since the higher the mass of the clump, the larger the part of the 
SUSY parameter space
that lies above the minimum detectable cross section, ${\langle \sigma v
\rangle}_{cont.}$, versus
$m_{\chi}$ curve;
equivalently, the higher the mass, the stronger the constraints on 
the SUSY parameter
space that the {\it non}-detectability of an otherwise detected clump
(via e.g. its synchrotron
emission~\cite{bot02}) imposes, given that if the clump is not 
detected by  ACTs, all the models
above the $\langle \sigma v \rangle_{cont.}$-$m_{\chi}$ curve will be excluded.

\begin{figure}
\epsfig{file=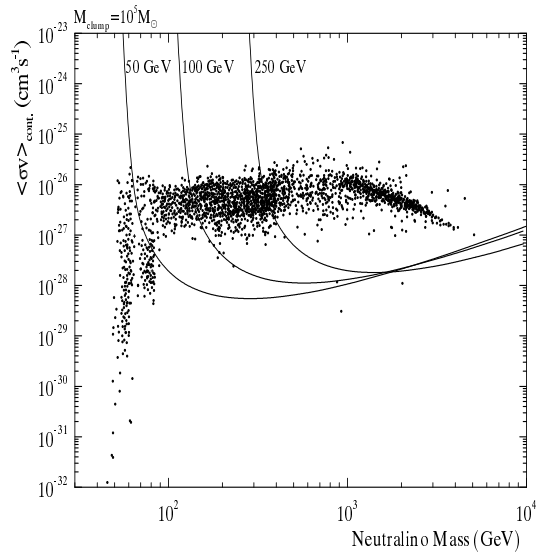, width=8.6cm, height=8.6cm}
\caption{\label{threshold} The minimum detectable $\langle \sigma v
\rangle_{cont.}$  versus
$m_{\chi}$ for a $10^{5} M_{\odot}$  Moore et al. clump, and
for three different energy thresholds, $E_{th}$:
50 GeV, 100 GeV and 250 GeV.
The lines reperesent
the 5-$\sigma$  detection for 100 hrs of observation, for an angular
resolution
$\sigma_{\theta}=6^{'}$, $5^{'}$ and
$3^{'}$, for energy threshold
$E_{th}=$ 50 GeV, 100 GeV and 250 GeV, respectively,
and for
an effective  collective area, $A_{eff}$, equal to $10^{8}$ cm$^{2}$
for the cases of  50 GeV and 100 GeV, and equal to
$3 \times 10^{8}$ cm$^{2}$  for the case of 250 GeV.
The other parameters are fixed as in FIG.\ref{conti-susy}.
}
\end{figure}

The minimum neutralino mass that can be explored also depends on the
observation features, and more importantly,  on the energy threshold.
To show the role of the energy  threshold, $E_{th}$, in the detectability of
the clumps for the SUSY  parameter space,
in FIG.\ref{threshold} we plot the same kind of curves as in
previous figures,
using  a $10^{5}M_{\odot}$ Moore et al. clump and for three different
energy thresholds: 50, 100 and 250 GeV, for 100 hrs of observation  and for
$A_{eff}$ and $\sigma_{\theta}$ in agreement  with
Table \ref{veritaspecs}.  Using a higher energy threshold means both a
larger $A_{eff}$ and a better angular resolution, as well as lower
contributions from the backgrounds, which makes it clear that ACTs are
well-suited
to explore the higher $m_{\chi}$ range of the neutralino parameter
space.  On the other hand,
the lower the energy  threshold the larger the
part of the parameter space that
ACTs can explore, but  even in the best case  of the smallest energy
threshold, the low mass parameter space which corresponds to a
broad range of  $\langle \sigma v \rangle_{cont.}$  will remain
unexplored regardless of the mass of the clump. This demonstrates the need
for complementary  observations at lower energies,  such as observations
with GLAST \cite{glast}, accelerator searches, and direct searches.

\begin{figure}
\epsfig{file=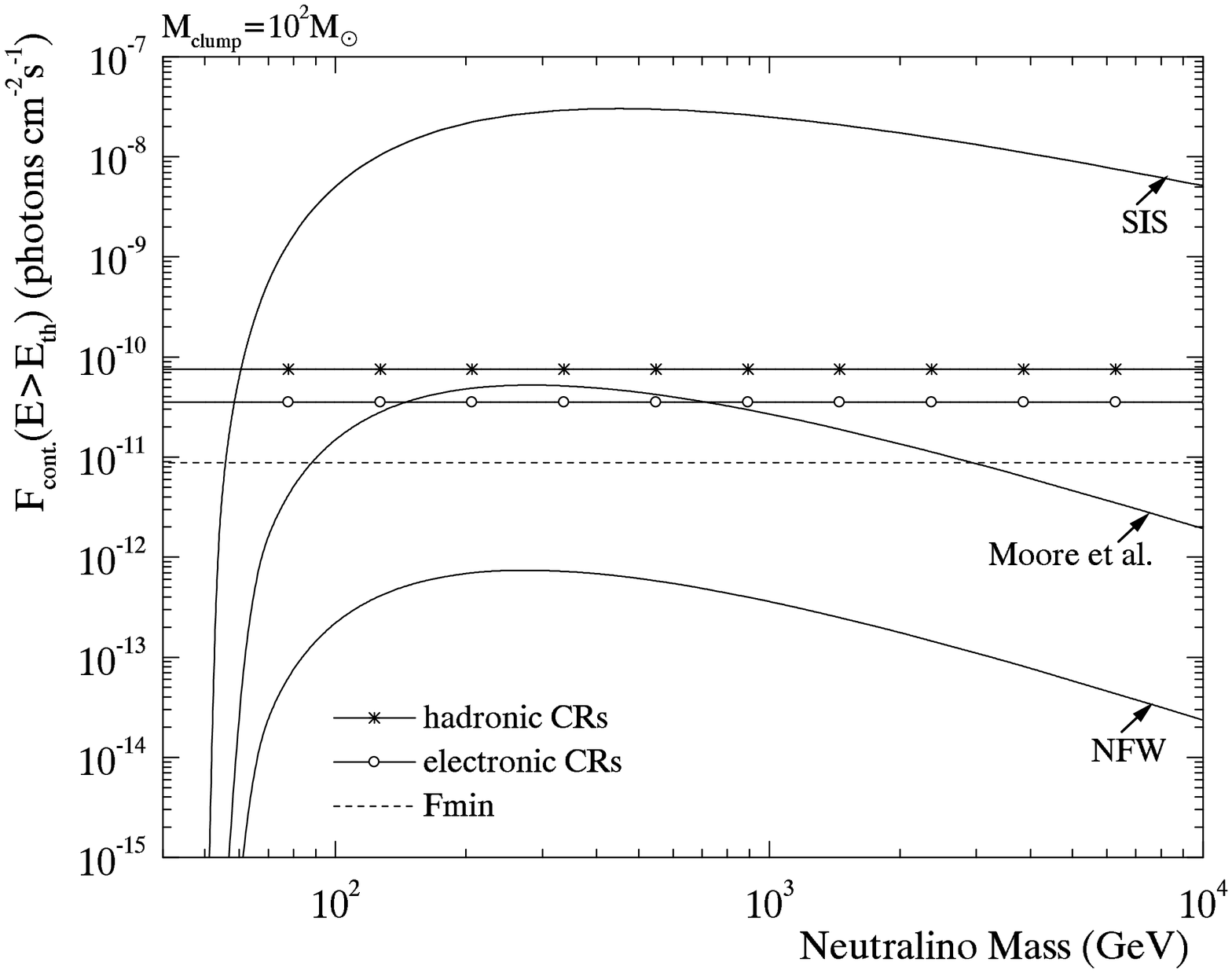, width=8.6cm, height=8.6cm}
\epsfig{file=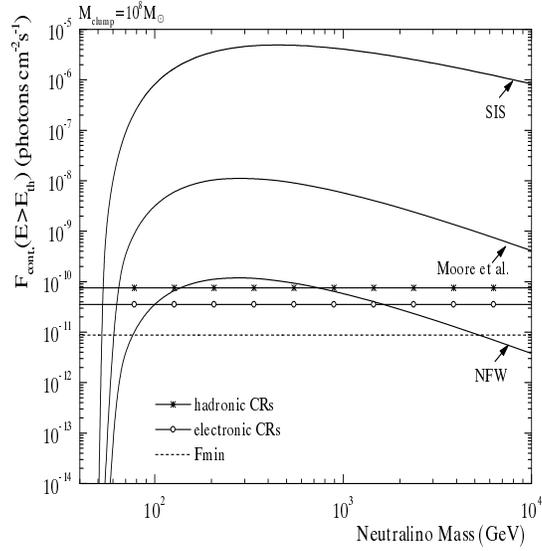, width=8.6cm, height=8.6cm}
\caption{\label{conti-flux}
The flux of   continuum
$\gamma$-rays as a function of $m_{\chi}$, for
    $\langle \sigma v \rangle_{cont.} = 5 \times
10^{-27}$ cm$^{3}$s$^{-1}$, and for the SIS, Moore et al., and NFW profile.
The clump masses used are $10^{2} M_{\odot}$
and $10^{8} M_{\odot}$.
Also shown are the two dominant background contributions,
the hadronic cosmic ray and the electronic cosmic ray induced
ones. The dashed line  represents
the minimum flux of the clump at the Earth, required so
that a 5-$\sigma$  detection level be achieved for
    $A_{eff}=10^{8}$ cm$^{2}$,  $E_{th}$=50 GeV,
$\sigma_{\theta}=6^{'}$, and 100 hrs of observation.
For $m_{\chi}$ values corresponding to fluxes  higher than $F_{min}$,
the clumps  will
be detectable in the continuum.
}
\end{figure}

Our second set of results concerns the flux coming from a clump as
a function of neutralino mass. Following the trend given by the DarkSUSY
models,  we choose a typical cross section  for the
annihilation into the continuum $\langle \sigma v\rangle_{typ} \simeq 5
\times 10^{-27}$ cm$^{3}$ s$^{-1}$. Note that the SUSY parameter space is
not extremely wide in the continuum case,
at least  for neutralino masses above
$\simeq$ 80 GeV, and thus the choice  of
a typical cross section is reasonable. The flux as a function of $m_{\chi}$
for $\langle \sigma v\rangle_{typ}$, and
for clump masses $M_{clump} =10^{2}M_{\odot}$
and $M_{clump} = 10^{8}M_{\odot}$  is shown
in FIG.\ref{conti-flux}. As above, we show the results for the three
different density profiles we used.
The energy threshold was set to $50$ GeV.
On the same plot and for the same energy threshold, the electronic
and hadronic backgrounds, as well as the
5-$\sigma$ minimum detectable
flux, $F_{min}$ (defined in Eq.(\ref{minflux})),
are plotted using $A_{eff}=10^{8}$
cm$^{2}$ and $t=3.5 \times 10^{5}$ s $(100$ hrs$)$.
In these plots a clump of a specific mass and density profile will be
detectable  for those neutralino masses for which its flux
is greater or equal to the minimum detectable flux $F_{min}$.
We will call this range of neutralino masses {\it the $m_{\chi}$ detectability
range}.

From  FIG.\ref{conti-flux}, we see that a $10^{2}M_{\odot}$
NFW clump
will not be detectable (regardless of $m_{\chi}$),
an SIS clump of the same mass has an $m_{\chi}$
detectability range  $m_{\chi} \ga  55$
GeV, whereas if the clump density distribution is better
described  by a Moore et al. profile, the $m_{\chi}$ detectability
range will be $87$ GeV $\la m_{\chi} \la 2.9$ TeV, approximately. In 
the case of a
$10^{8} M_{\odot}$ clump, the SIS profile has an $m_{\chi}$ detectability range
$m_{\chi} \ga  53$ GeV. Similarly for the Moore et al. $10^{8}M_{\odot}$ clump
$m_{\chi} \ga 65$ GeV, whereas
the same clump if modeled with an NFW profile will have an $m_{\chi}$ 
detectability
range  $77$ GeV $\la  m_{\chi} \la 5$ TeV.
Focusing on the intermediate case, the Moore et al. profile, we see that
in terms of constraints on $m_{\chi}$, the detectability of low mass clumps
provide both a lower and an upper limit (i.e., detectability range of the  form
$m_{\chi,min} \la m_{\chi} \la m_{\chi,max}$), whereas the higher-mass clump
detectability provides only a weaker lower limit (i.e., detectability range of
the form $m_{\chi} \ga m_{\chi,min'}$).  Note that the same 
conclusions if applied in
the case of a clump known to exist  but which turns out to be 
non-detectable by ACTs,
imply  complementary  to the above limits for the neutralino 
parameter space. Namely,
the  non-detectability of the clump will imply that $m_{\chi}$ lies 
in the range : all
$m_{\chi}$ values predicted by SUSY $-$ detectability range. 
Actually, as can be seen
from all figures referring to the continuum signal, it will be the 
non-detectability that
will impose the strongest constraints on the SUSY parameter space. 
Furthermore, although
the larger the clump mass the easier to detect, having a range of 
clump masses is needed
in order to determine the neutralino parameters. As we discuss below, the
constraints can become stronger  when information on the line
emission
is combined with the better detectable continuum $\gamma$-rays.

Lastly, given the SUSY parameter space and the $\langle \sigma
v \rangle_{cont.}/ m_{\chi}^{2}$ dependence of the flux, we find the
minimum $F_{min}$ and maximum $F_{max}$ flux
above $50$ GeV that a clump of a specific mass
and density profile can emit. We do that
by finding from the SUSY parameter space what $\langle \sigma
v \rangle_{cont.}$ and $m_{\chi}^{2}$ combinations yield the minimum and
maximum $\langle \sigma v \rangle_{cont.} /m_{\chi}^{2}$  ratio,
respectively.
In this way,
we bracket the flux of a clump and we know that, even for density
profiles other than the three ones used here, as long as they are somewhere
between the NFW and the SIS, their flux will be somewhere in the interval
$[F_{min}, F_{max}]$. The results of this calculation are shown
in Table \ref{fluxes}. For the minimum
possible clump flux, $F_{min}$, we
chose   $m_{\chi}=5$ TeV and $\langle \sigma v \rangle= 10^{-27}$
cm$^{3}$ s$^{-1}$. This combination
   yields a representative lower  flux value, at least for the
   horizontal feature  of  SUSY models in the large $m_{\chi}$ range (even
lower fluxes are obtained for the low mass and low cross section part of
the SUSY parameter space, but this range is not likely
to be accessible by  ACTs, anyway).
    For the maximum possible clump flux, $F_{max}$,
    we used $m_{\chi}=60$ GeV and
$\langle \sigma v \rangle=2 \times 10^{-26}$ cm$^{3}$ s$^{-1}$,  which is
a combination that quite reliably gives the maximum possible flux.
For comparison, in Table \ref{min}
we give the minimum
required flux, $F_{min,det.}$, so that a 5-$\sigma$ detection be 
obtained for  certain
$E_{th}, A_{eff}$ and $t$ combinations, as was calculated using Eq. 
(\ref{minflux}) and
assuming the hadronic and electronic backgrounds as the dominant ones.

\begin{table}
\caption{\label{fluxes} Minimum and maximum possible fluxes for the three
different density profiles -- see text for details.
All  fluxes are in photons cm$^{-2}$s$^{-1}$
and masses are in $M_{\odot}$.}
\begin{tabular}{lccccc}  \hline \hline
$M_{clump}$ & & $10^{2}$ & $10^{4}$ & $10^{6}$ & $10^{8}$ \\ \hline
    & $F_{min}$  &  $2.1 \times 10^{-9}$ &  $1.3 \times 10^{-8}$
&  $7.6 \times 10^{-11}$    &   $3.4 \times 10^{-7}$ \\
\rb{SIS} & $F_{max}$ & $2.8 \times 10^{-8}$  & $1.7 \times 10^{-7}$
&  $9.7 \times 10^{-10}$  &  $4.4 \times 10^{-6}$  \\  \hline
& $F_{min}$ &  $6.5 \times 10^{-13}$ & $4.0 \times 10^{-12}$ &
$2.4 \times 10^{-11}$ &  $1.4 \times 10^{-10}$   \\
\rb{Moore} & $F_{max}$ &  $1.6 \times 10^{-10}$
& $9.7 \times 10^{-10}$ &
$6.1  \times 10^{-9}$  &  $3.3 \times 10^{-8}$ \\  \hline
& $F_{min}$ & $7.9 \times 10^{-15}$ & $4.3 \times 10^{-14}$  &
$2.4 \times 10^{-13}$  &  $1.3 \times 10^{-12}$   \\
\rb{NFW} & $F_{max}$ & $2.6 \times 10^{-12}$  & $1.4 \times 10^{-11}$
&   $8.1 \times 10^{-11}$   &    $4.2 \times 10^{-10}$   \\ \hline \hline
\end{tabular}
\end{table}

\begin{table}
\caption{\label{min} Minimum detectable fluxes for specific energy
thresholds, effective areas and integration times so that a 5-$\sigma$
detection be obtained. All  fluxes are in photons cm$^{-2}$s$^{-1}$,
energies in GeV, areas in cm$^{2}$ and
integration times in s.}
\begin{tabular}{lccc}  \hline \hline
$E_{th}$ & $A_{eff}$ & t & $F_{min,det.}$ \\   \hline
50 & $1.0 \times 10^{8}$ & $3.5 \times 10^{5}$ & $8.9 \times 10^{-12}$ \\
100 & $1.0 \times 10^{8}$ & $3.5 \times 10^{5}$ & $4.6 \times 10^{-12}$ \\
250 & $1.0 \times 10^{8}$ &  $3.5 \times 10^{5}$ & $2.0 \times 10^{-12}$\\
50 & $1.0 \times 10^{9}$ & $3.5 \times 10^{5}$ & $2.8 \times 10^{-12}$ \\
50 & $1.0 \times 10^{8}$ & $ 2.6 \times 10^{6}$ & $3.2 \times 10^{-12}$ \\
50 & $1.0 \times 10^{8}$ & $3.2 \times 10^{7}$ & $1.1 \times 10^{-13}$ \\
\hline \hline
\end{tabular}
\end{table}

From the two tables we see that the least likely
case of  clumps, namely clumps  as dense as the
SIS, can be easily detectable in less than 100 hrs. In the case of
the Moore et al. profile,
the higher mass clumps  ($\geq 10^{6} M_{\odot}$)
will be easily detectable since both $F_{min}$  and $F_{max}$  exceed the
minimum required flux, $F_{min,det.}$, to achieve a 5-$\sigma$ detection level.
The lower mass clumps have an easily detectable $F_{max}$, but not
necessarily a detectable $F_{min}$.  A $10^{4} M_{\odot}$ Moore
et al. clump would
have a detectable $F_{min}$  for either a large integration times ($2.6
\times 10^{6}$ s $\simeq $1 month ), or for larger collective areas
(e.g., $10^{9} {\rm cm}^{2}$ at 50 GeV), which is not among
the abilities of both
existing and upcoming ACTs.
Finally, NFW clumps with $M_{clump} \ga 10^{6} M_{\odot}$  are detectable
and this is particularly important given the EGRET constraints placed on the
most concentrated clumps, such as SIS and Moore et al. ones.
In contrast, even a whole
year of observation ($3.2 \times 10^{7}$s)
would not be enough to detect smaller
mass NFW clumps with fluxes that are closer to the corresponding
$F_{min}$.

\subsection{Monochromatic $\gamma$-rays}

\begin{figure}
\epsfig{file=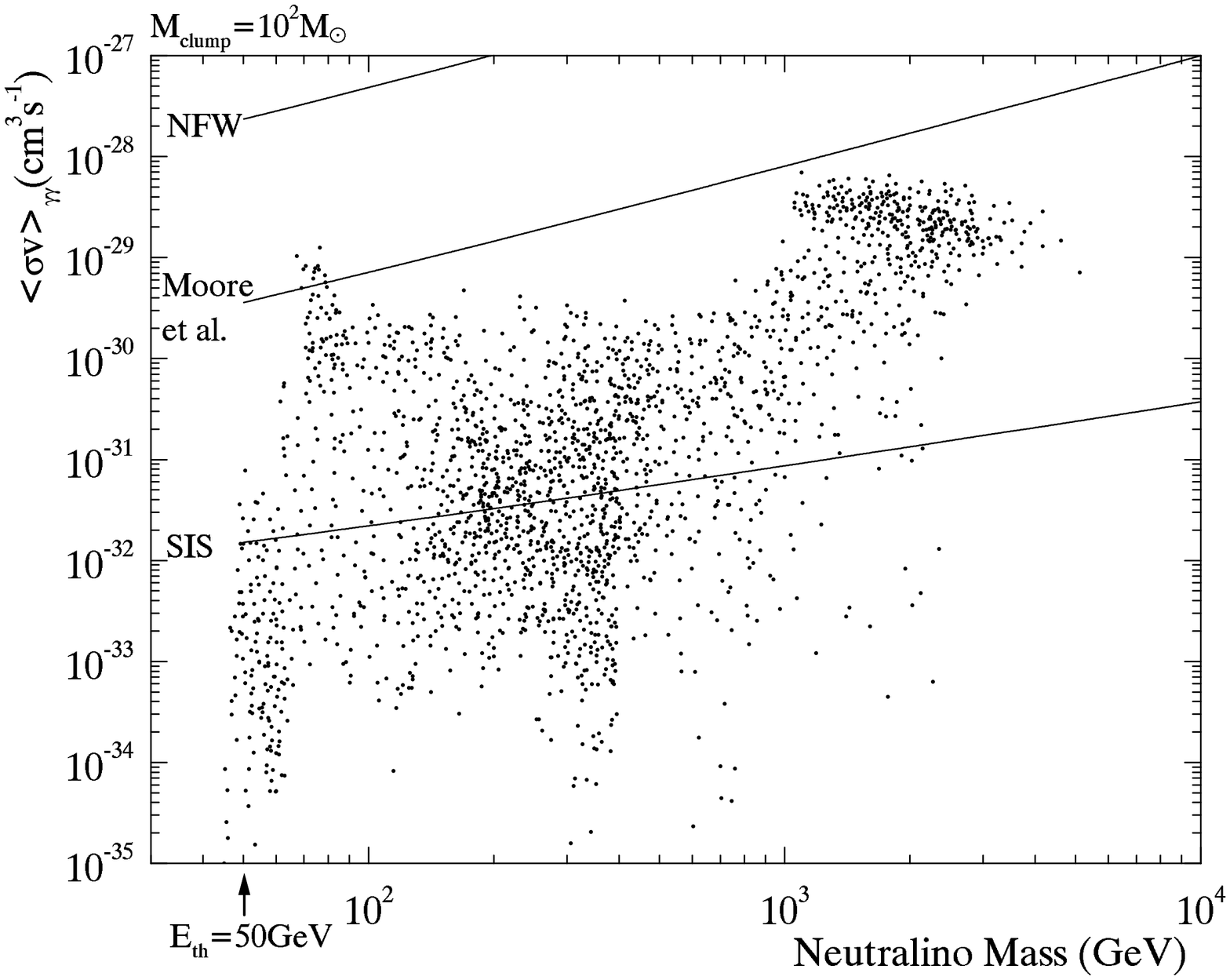, width=8.3cm, height=8.3cm}
\epsfig{file=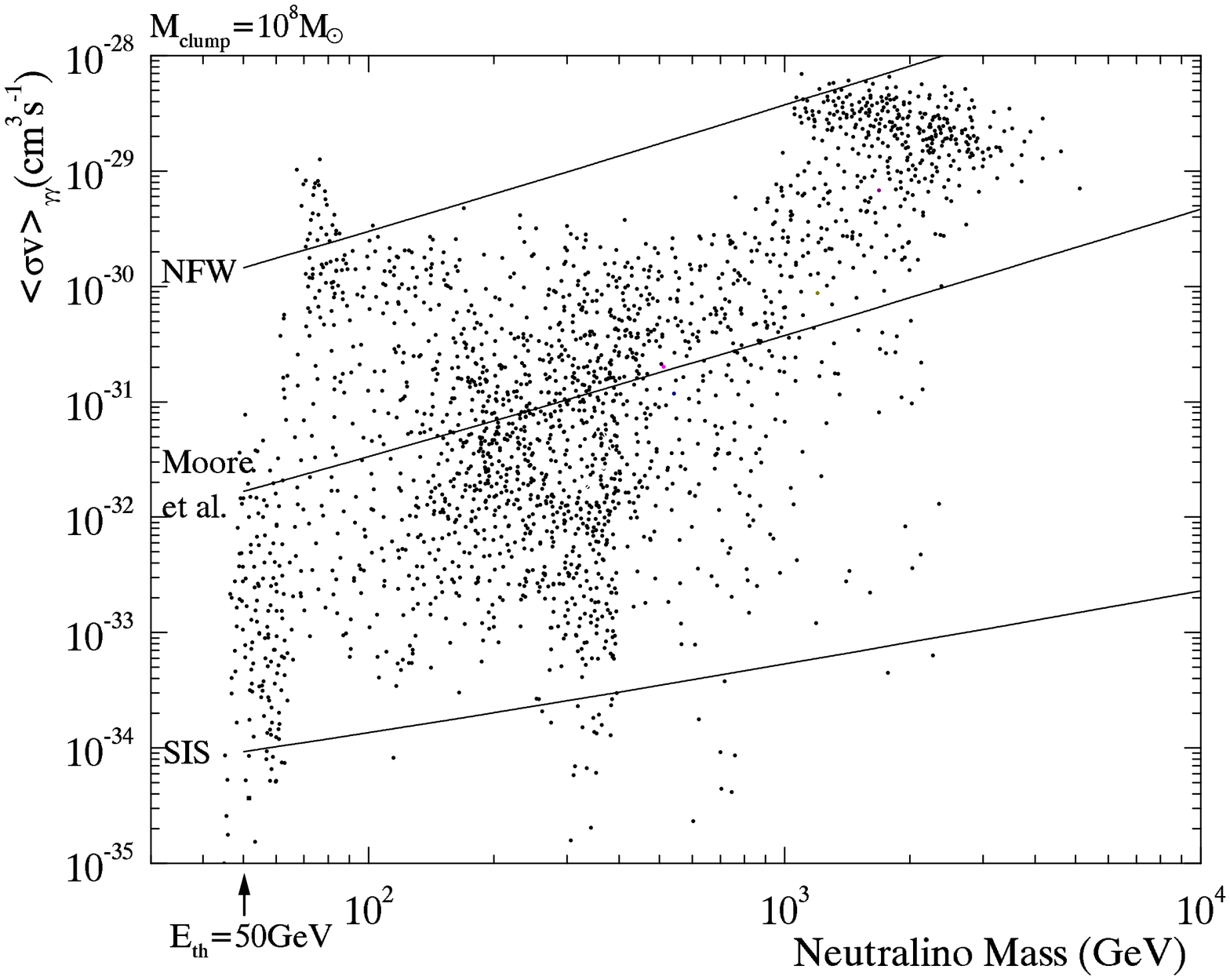, width=8.3cm, height=8.3cm}
\caption{\label{2g-susy}  The minimum detectable $\langle \sigma v
\rangle_{\gamma\gamma}$ as a
function of $m_{\chi}$ for  the SIS, the Moore et al., and the NFW
profile. The clump masses used are
$10^{2} M_{\odot}$ and
$10^{8} M_{\odot}$.  The dots represent
allowed SUSY models. We require a 5-$\sigma$   detection
level for  $A_{eff}=10^{8}$ cm$^{2}$,  $E_{th}$=50 GeV,
$\sigma_{\theta}=6^{'}$, energy resolution
$R_{E}=15 \%$, and for 100 hrs of observation.
For a specific  clump mass and a specific density profile, and for
the above instrument and observation characteristics, only the
SUSY models that lie above the corresponding curve will yield
a detectable signal.
}
\end{figure}

The minimum detectable cross section
$\langle \sigma v\rangle_{\gamma\gamma}$  for
annihilation into monochromatic $\gamma$-rays with energy equal to
$m_{\chi}$, as a function of $m_{\chi}$ is shown for
$M_{clump} =10^{2}M_{\odot}$  and
$M_{clump} = 10^{8}M_{\odot}$
in  FIG.\ref{2g-susy}.
The constraining curves stop at $E_{th}$, which in
this case is 50 GeV (below this the cross section is non-zero, but the
number of photons with energy above 50 GeV coming from neutralinos with
mass less than 50 GeV is, of course, 0).  A first, clear difference,
in comparison with
   the continuum $\gamma$-rays, is the fact that the SUSY parameter
space is significantly wider with respect to the cross section. Therefore,
imposing constraints on the SUSY parameter space using the detectability
(or non-detectability) of the clumps through their  $\gamma\gamma$ 
line emission
will be  harder than through their continuum emission.
Nevertheless, if lines are  detected in  addition to the
continuum, the degeneracy in the SUSY
parameter space will be lifted since
a narrower range of models would fit both
detections.

\begin{figure}
\centering
\epsfig{file=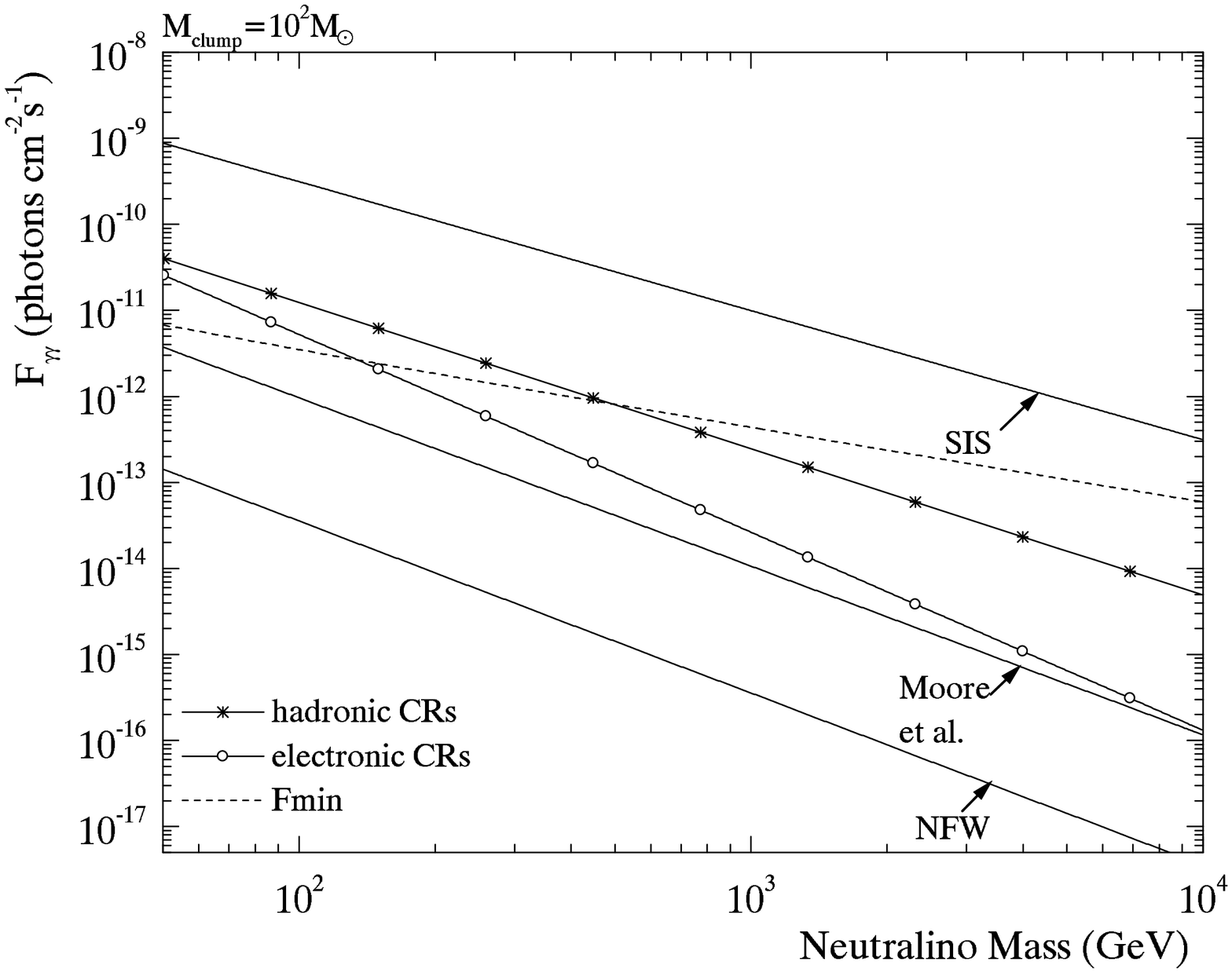, width=8.6cm, height=8.6cm}
\epsfig{file=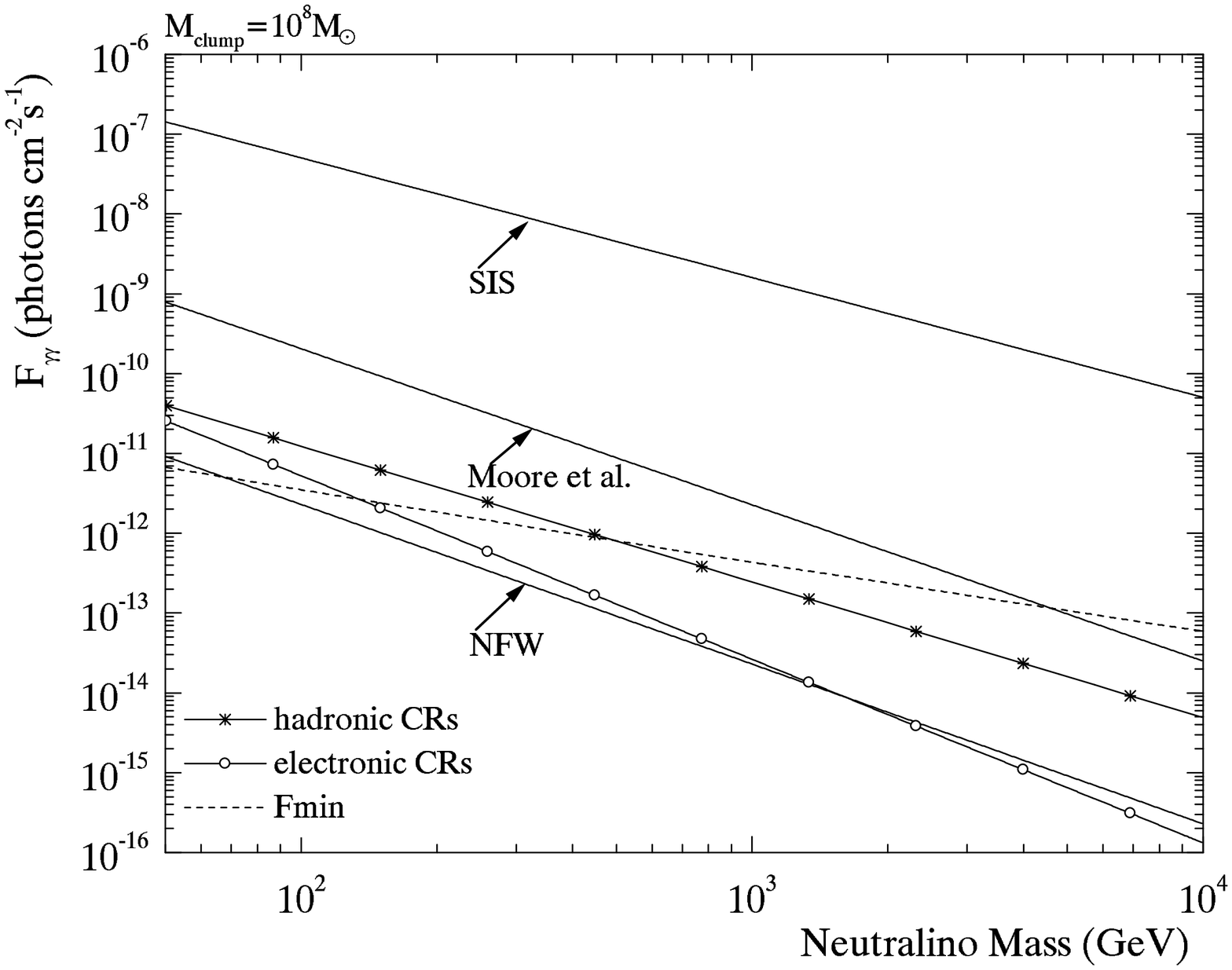, width=8.6cm, height=8.6cm}
\caption{\label{2g-flux}  The flux of $\gamma\gamma$ monochromatic
$\gamma$-rays as a function of $m_{\chi}$,
assuming $\langle \sigma v \rangle_{\gamma\gamma} \simeq 2 \times
10^{-30}$ cm$^{3}$s$^{-1}$, and for the SIS, the Moore et al., and
the NFW profile.
The clump masses used are $10^{2} M_{\odot}$ and
$10^{8} M_{\odot}$.
Also shown as
   functions of $E_{\gamma}=m_{\chi}$,
are the two dominant background contributions,
the hadronic cosmic ray and the electronic cosmic ray induced
ones. The dashed line represents the minimum flux, $F_{min}$, required
so that a 5-$\sigma$   detection be achieved for
$A_{eff}=10^{8}$ cm$^{2}$,  $E_{th}$=50 GeV,
$\sigma_{\theta}=6^{'}$, and 100 hrs of observation.
For $m_{\chi}$ values corresponding to fluxes  higher than $F_{min}$,
the clumps will
be detectable in the $\gamma\gamma$ line.
}
\end{figure}

This is depicted  more quantitatively in FIG.\ref{2g-flux} where
we present the flux  of monochromatic $\gamma$-rays
as a function of $m_{\chi}$, as well as the minimum detectable
flux and the two major background contributions.
Note that in
this case both the minimum detectable flux and the background
fluxes depend on $m_{\chi}$, given that the
energy of the monochromatic $\gamma$-rays is equal to $m_{\chi}$.
As we did in the case of the continuum,
to find the flux coming from a given clump
we assume a typical cross section, in order to express the flux only as
a function of $m_{\chi}$. Based on the SUSY parameter space  presented
in FIG.\ref{2g-susy}, we choose the large cross section region and use
$\langle \sigma v\rangle_{\gamma\gamma}=2 \times 10^{-30}$ cm$^{3}$s$^{-1}$.
This value yields a flux close to the maximum possible flux
   such that a comparison can be
attempted with the continuum case, which is generally stronger.
We see, however, from FIG.\ref{2g-flux} that even in this case,  the
$\gamma\gamma$
line will not be observable with upcoming ACTs. More specifically,    for
$M_{clump} = 10^{2}M_{\odot}$
the chances of observing such a light clump  in
the monochromatic $\gamma$-ray lines are bad, regardless of neutralino mass.
With respect to the density profile, neither the Moore et al. nor the
NFW will be detectable. In the case of a density profile between the Moore
et al. and the SIS,   observability could be achieved imposing at the same
time an upper limit on $m_{\chi}$. In the case of  $M_{clump} =  10^{8}
M_{\odot}$,
a Moore et al. clump will be detectable as long as $m_{\chi} \la$ 4.5
TeV,  whereas the same clump, if described by an NFW profile,
would be detectable only if $m_{\chi} \la $68 GeV.

\begin{figure}
\centering
\epsfig{file=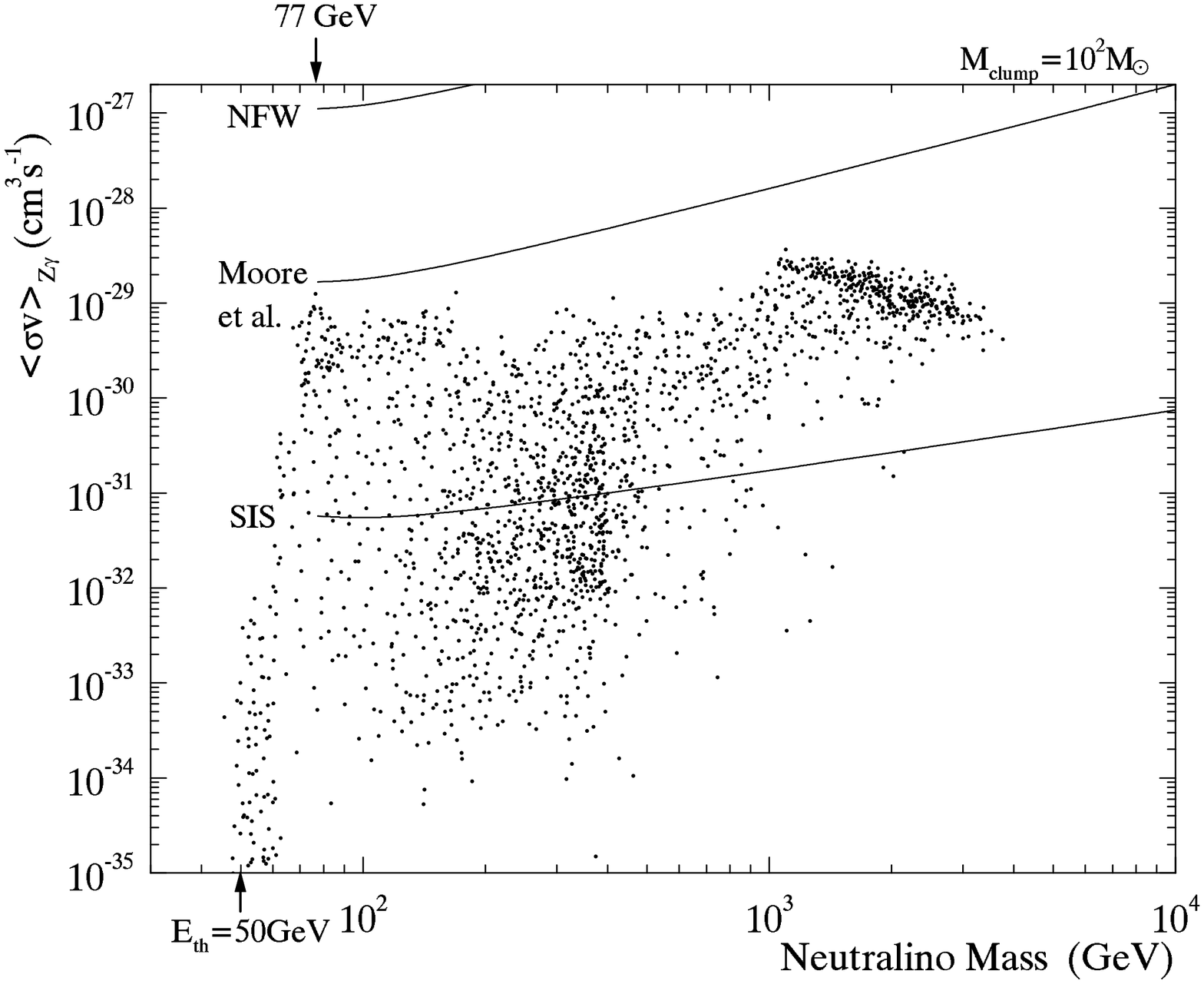, width=8.6cm, height=8.6cm}
\epsfig{file=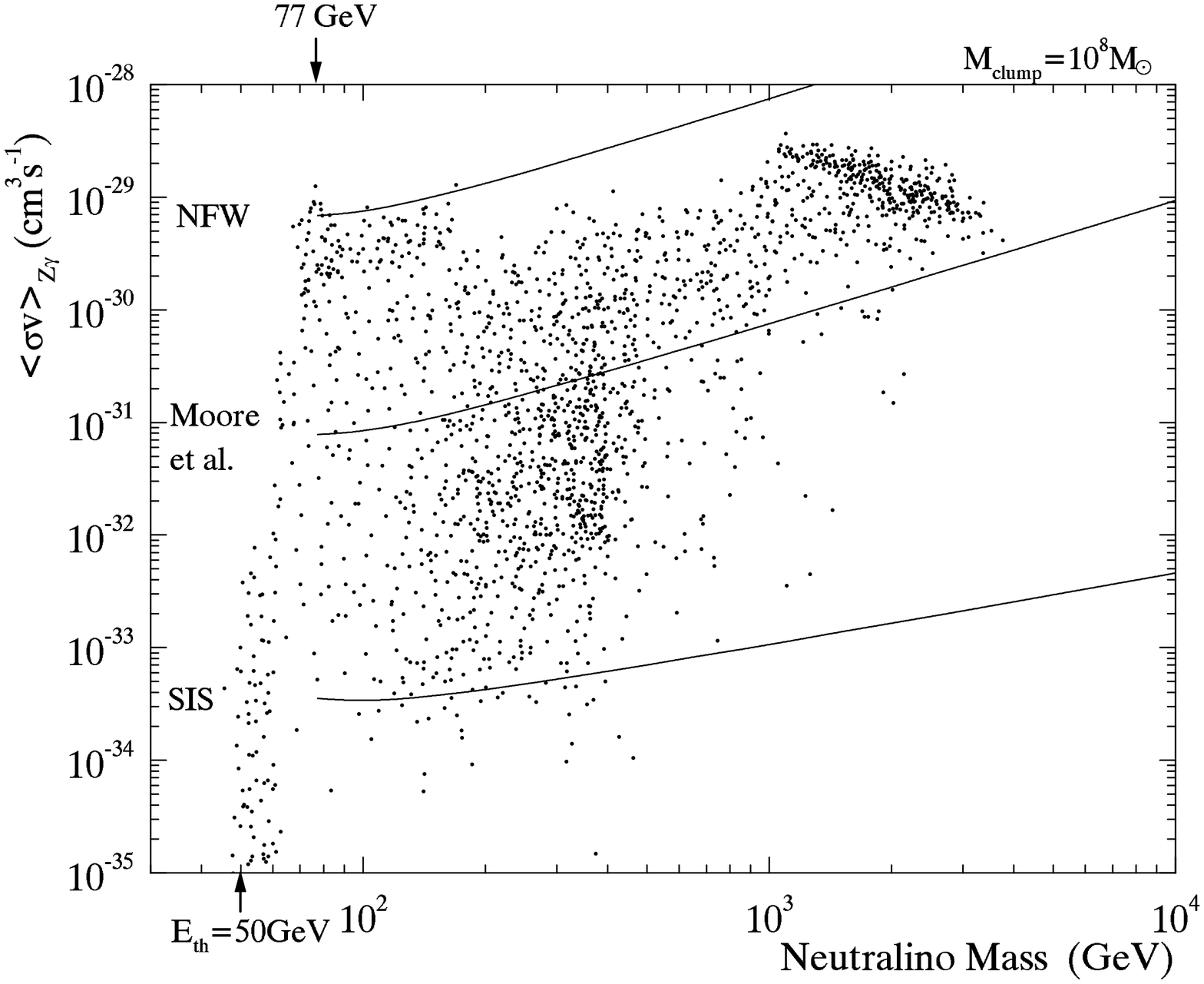, width=8.6cm, height=8.6cm}
\caption{\label{2z-susy}  The minimum detectable $\langle \sigma v
\rangle_{Z\gamma}$ as a
function of $m_{\chi}$ for  the SIS, the Moore et al., and the NFW
profile. The clump masses used are
$10^{2} M_{\odot}$ and
$10^{8} M_{\odot}$.  The dots represent
allowed SUSY models. We require a 5-$\sigma$   detection
level for  $A_{eff}=10^{8}$ cm$^{2}$,  $E_{th}$=50 GeV,
$\sigma_{\theta}=6^{'}$, energy resolution
$R_{E}=15 \%$, and for 100 hrs of observation.
For a specific  clump mass and a specific density profile, and for
the above instrument and observation characteristics, only the
SUSY models that lie above the corresponding curve will yield
a detectable signal.
}
\end{figure}

The case is similar for the $Z\gamma$ line.  Results for the SUSY
parameter space are presented in FIG.\ref{2z-susy}, for $M_{clump}
= 10^{2}M_{\odot}$ and $10^{8}M_{\odot}$.
In these figures we require that $E_{\gamma} \geq E_{th}$.
Using  the fact that $E_{\gamma}=m_{\chi} -{m_{Z}^{2}/ 4 m_{\chi}}$,
this implies that the neutralino mass
\begin{equation}
m_{\chi} \geq \frac{E_{th} + \sqrt{m_{Z}^{2} + E_{th}^{2}}}{2}
\end{equation}
which translates to $m_{\chi} \ga 77 $ GeV for $E_{th}= 50$ GeV.
The two  lines are expected to coincide at relatively large $m_{\chi}$,
and
thus, the conclusions for both are similar with the difference that due to
the above constraint the $Z\gamma$ line
explores the parameter space starting from higher neutralino masses
compared
to the $\gamma \gamma$ line. In addition, the SUSY models
corresponding to $m_{\chi}$  above 500 GeV  are
less dispersed
in the case of the   $Z\gamma$ line than  in the case of
the $\gamma\gamma$.
Again, this will help lift degeneracies of neutralino models,
if any clumps are detected.

\begin{figure}
\centering
\epsfig{file=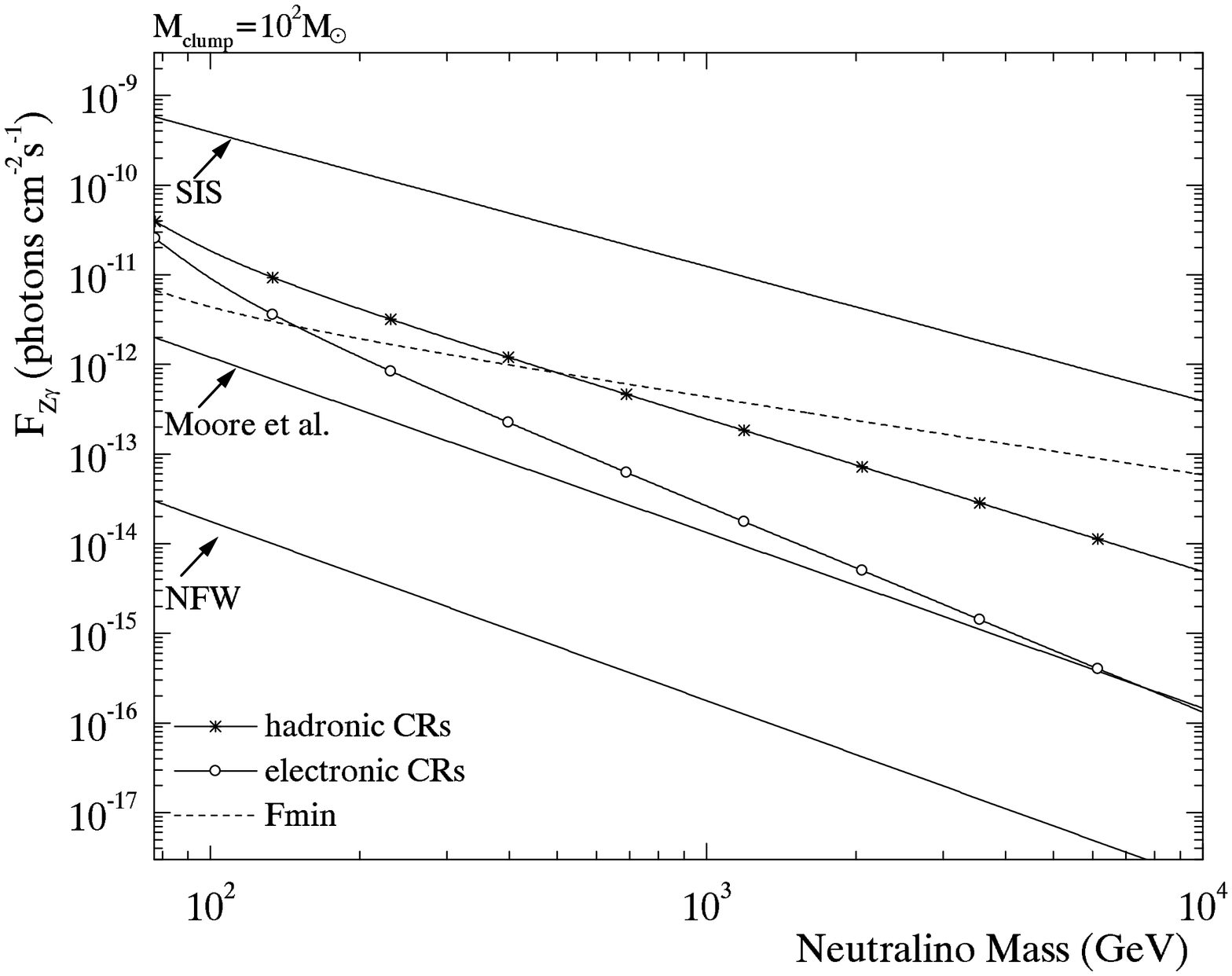, width=8.6cm, height=8.6cm}
\epsfig{file=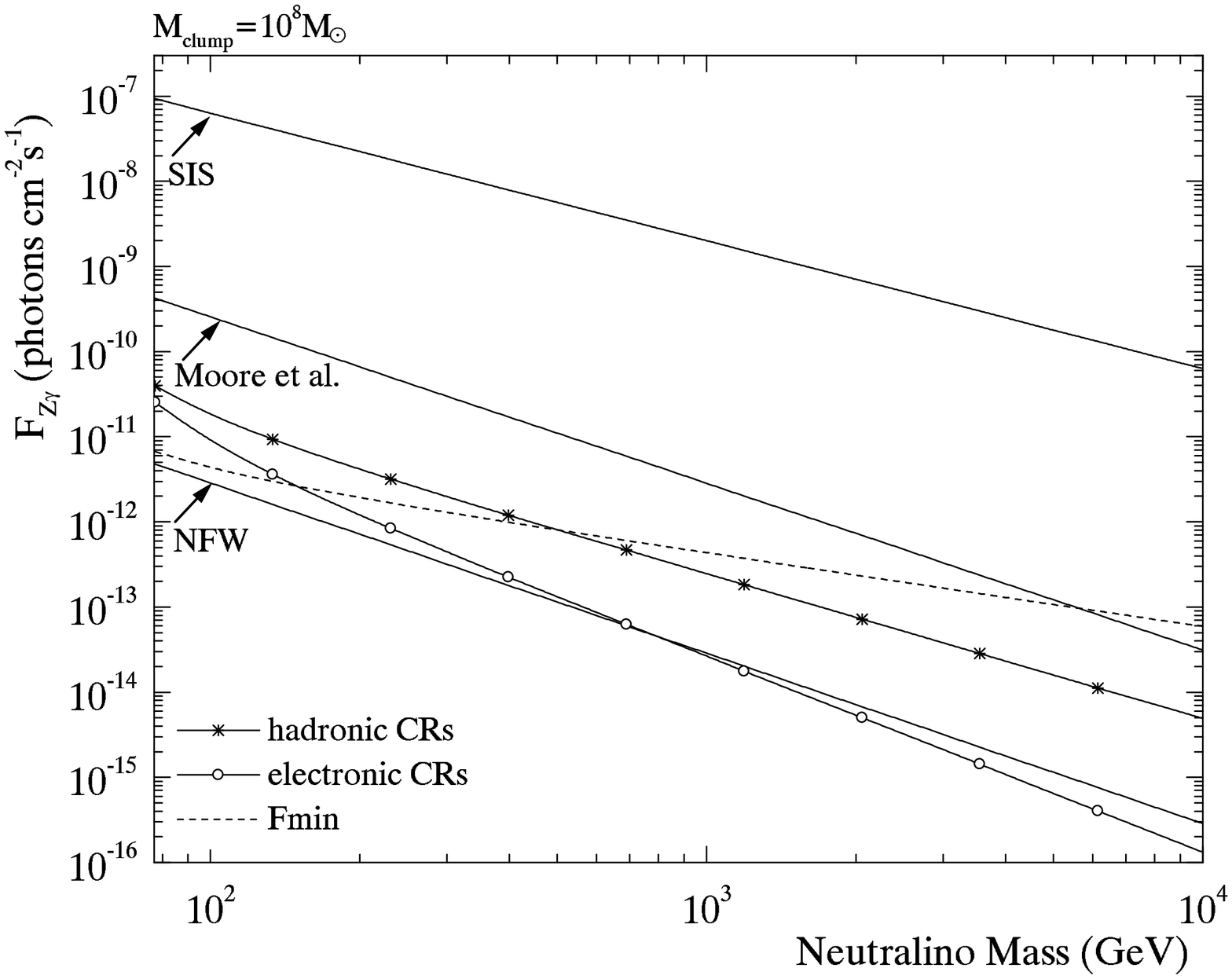, width=8.6cm, height=8.6cm}
\caption{\label{2z-flux}
The flux of $Z\gamma$ monochromatic
$\gamma$-rays as a function of $m_{\chi}$,
assuming $\langle \sigma v \rangle_{Z\gamma} \simeq 3 \times
10^{-30}$ cm$^{3}$s$^{-1}$, and for the SIS, the Moore et al., and
the NFW profile.
The clump masses used are $10^{2} M_{\odot}$ and
$10^{8} M_{\odot}$.
Also shown, as functions of $m_{\chi}$,
   are the two dominant background contributions,
   the hadronic cosmic ray and the electronic cosmic ray induced
ones. The dashed line represents the minimum flux, $F_{min}$, required
  so that 5-$\sigma$   detection
be achieved for
$A_{eff}=10^{8}$ cm$^{2}$,  $E_{th}$=50 GeV,
$\sigma_{\theta}=6^{'}$, and 100 hrs of observation.
For $m_{\chi}$ values corresponding to  fluxes  higher than $F_{min}$,
the clumps  will
be detectable in the $Z\gamma$ line.
}
\end{figure}

The flux as a function of neutralino mass assuming  a
maximum $\langle \sigma v\rangle_{Z\gamma} = 3 \times 10^{-30}$
cm$^{3}$ s$^{-1}$, is shown in FIG.\ref{2z-flux}.
In general the
conclusions are the same as in the $\gamma \gamma$ line. The light clumps
are
not very
promising with respect to their detectability.  A
$10^{8} M_{\odot}$ Moore  et al. clump will be detectable   as long as
$m_{\chi} \la $ 6 TeV. Note, however, that this is true under the assumption
of
  the maximum  $\langle \sigma v\rangle_{Z\gamma} $, and thus of
the maximum possible flux.

\section{Discussion and Conclusions}

The results presented above were based on specific choices for the
detector and the observation parameters.
It is worth noting that these results can be
easily scaled with some of these parameters. For instance, the
effective area $A_{eff}$ and observation time $t$ have a simple
scaling.  The minimum detectable  $\langle \sigma v
\rangle$ for a  specific $m_{\chi}$ scales  as $ 1/ (A_{eff}
t)^{1/2}$. An order of magnitude lower $\langle \sigma v
\rangle$ could be explored using two orders of magnitude higher
$A_{eff} t$, for example, $A_{eff}=10^{9}$ cm$^{2}$ and $t=3.5 \times
10^{6}$ s ($\simeq$ 40 days). This would be very helpful in the case
of the hard to detect lines (see FIG.\ref{2g-susy} and
FIG.\ref{2z-susy}). If some clumps are detected in the continuum, a
reasonable strategy would be to subsequently search for the weaker
line signal over a longer $t$.
Similarly, the minimum detectable flux
plotted in FIGs.\ref{conti-flux}, \ref{2g-flux}, and \ref{2z-flux},
depends on
the combination $A_{eff}t$.
With respect to the angular resolution $\sigma_{\theta}$,
the minimum
detectable  $\langle \sigma v \rangle$ has a dependance of the form
$\propto 1/ \sqrt{\Delta \Omega}$, approximately, while the  minimum
detectable fluxes are $\propto \sqrt{\Delta \Omega}$.
Note that although the nearest clumps  are resolved,
we used the angular resolution of the instrument and not the
angular size of the clumps as the relevant angle for the collection
of  background noise, assuming that we
observe the central pixel of
the source -- where most of the source counts originate from -- treating
thus, the source with respect to the backgrounds
as being effectively unresolved.
  If, instead, one uses the angular size of the
object as the relevant angle for the collection of noise, both the
detectable cross sections and the minimum detectable fluxes are about
two and one orders of magnitude higher for a $10^{8} M_{\odot}$ and a
$10^{2}M_{\odot}$ clump, respectively.

An essential parameter in our calculations is the distance of the
source from the Earth, which we assumed to be the estimated  distance to
the nearest clump. However, our results can be easily extended  to
larger  distances or, equivalently, to lower numbers of
clumps  given the $1 / d^{2}$ dependence of the flux,
which is actually the most significant dependence of our
results on the distance~\cite{three}.
For example, one order of magnitude larger
distance makes the minimum $\langle \sigma v \rangle$ that can be
explored two orders of magnitude higher; the same
is true for the minimum detectable fluxes.
This is very useful given the
recent bounds on SIS and Moore et al. clumps from EGRET data~\cite{abo02}.
More specifically, only NFW clumps
can populate the whole halo as seen in CDM simulations without overwhelming
EGRET limits. For the other density profiles, consistency with
EGRET data could imply, e.g., a smaller than currently believed
clump abundance, which would translate, in our case, to larger distances
of the clumps. Using the above scaling of our results with distance, it is
easy to conclude with respect to the clump detectability  in any
other case that would change the distance estimate.
Similar is the case for the density profile assumed for the clumps;
the detectability of a density profile, e.g., with central
concentration in between that of the Moore at al. and the NFW profiles,
can be predicted, at least qualitatively, using our results.

The choice of the energy threshold  will be a crucial
factor determining what will be the minimum $m_{\chi}$ values
accessible  to  ACTs. An energy threshold low
enough, along with the ability of  ACTs
to identify primaries leading thus to backgrounds smaller by
an order of magnitude, are highly desirable for future
ACTs, if they are to be used to explore the SUSY parameter space and detect
even the less concentrated clumps such as the NFW case.

In addition, we would like to re-emphasize the strategy of
combining observations in the continuum and the  annihilation lines
for different clump masses in order to constrain more effectively the
SUSY parameter space. For example, from FIGs.\ref{conti-flux}
and~\ref{2g-flux} we saw that a low mass clump has an $m_{\chi}$
detectability range of the form $m_{min}\la m_{\chi} \la m_{max}$.
Combining this with line observations, one could narrow
the range of neutralino masses. More precisely, one should study
the allowed models in SUSY parameter space that fit the detected
continuum flux and check which ones give a detectable line signal.

The detectability of the lines can yield $m_{\chi}$ detectability
ranges of the form $m_{\chi} \leq
m_{max}^{'}$ for various combinations of clumps masses and
density profiles, assuming a typical or an optimum $\langle \sigma v
\rangle$ for the neutralino annihilation. Assuming for the lines the maximum
$\langle \sigma v\rangle$,
a $10^{8} M_{\odot}$ Moore et al. clump   will be
detectable both in the continuum and in the lines, if $m_{\chi}$ is
somewhere between 65 GeV and 4.5 TeV, where  $ m_{\chi}\ga$ 65 GeV
comes from the continuum limit and $m_{\chi} \la $4.5 TeV comes
from the lines. Although not a very narrow interval in $m_{\chi}$,
excluding the very low mass end of the allowed neutralino mass
range in combination with an upper $m_{\chi}$-limit for detectability,
can be very useful. It
would narrow down the range in $\langle \sigma v
\rangle_{cont.}$ significantly.
Another  example is  that of  a $10^{8} M_{\odot}$
NFW clump which is detectable in either the continuum ( detectability 
range: 77 GeV $\leq
m_{\chi} \la $ 5 TeV) or the $\gamma
\gamma$ line (detectability range: $m_{\chi} \leq$ 68 GeV), but not 
both. This fact can, in
principle, be used to extract information about the mass of the clump.

Our study of the detectability of clumps by ACTs is closely tied to
the  lack of knowledge with respect to the SUSY parameters.
Thus our results can be used in two ways, depending mainly on what will
be determined  first, the neutralino parameters or the distribution
and profiles of CDM clumps.  If clumps are discovered first (for
example, through synchrotron emission or from GLAST
$\gamma$-ray observations), $\gamma$-ray
studies by ACTs can help narrow the neutralino
parameter space. In most cases, the strongest constraints on the SUSY
parameter space arise from the
non-detectability in $\gamma$-rays of a clump
seen, e.g.,  in the microwaves~\cite{bot02} rather than
its detectability as, for example, in the case of a $10^{8} M_{\odot}$
Moore et al. clump. Such a clump will
be detectable in both the continuum and  the lines
if $m_{\chi}$ is  between 65 and 4.5 TeV. This large
range of neutralino masses can be ruled out if the clump is not detected
by VERITAS.
This can become quite useful in the future. For the time
being, what is even more important is that such a clump, if
existent, will
be easily detectable almost for any possible  neutralino
mass.

Summarizing, we conclude that the chances of detectability of dark matter
clumps in the galactic halo, due to continuum $\gamma$-rays
produced by neutralino annihilation via upcoming ACTs
are very good, especially in the case of relatively massive and
highly centrally concentrated  clumps. The signatures expected in the
form of the $\gamma\gamma$  and the $Z\gamma$ lines are less
easily detectable, but can be quite useful in lifting degeneracies
among neutralino SUSY parameters. If we figure out  the neutralino
parameters first, ACT searches will be strong tests of the
CDM model predictions with respect
both the substructure in galactic halos and
the  density profiles of the substructure clumps.
ACT searches will be able to explore large ranges of  neutralino
masses above $\ga$ 50 GeV (or more, depending on the
energy threshold). These studies will complement
the searches for lower mass neutralinos, accessible through high energy
accelerators, direct and other indirect detection methods, and satellite
$\gamma$-ray telescopes such as GLAST.

\section*{Acknowledgements}
We thank Paolo Gondolo for help with DarkSUSY and Roberto Aloisio,
Pasquale Blasi, and  Craig Tyler, for many interesting discussions.  This
work was supported in part by the NSF through grant AST-0071235 and DOE
grant
DE-FG0291-ER40606 at the University of Chicago and at the Center for
Cosmological Physics by grant NSF PHY-0114422.

\begin{thebibliography}{9}

\bibitem{jkg96}
   G. Jungman, M. Kamionkowski, K. Griest,   Phys.\ Rep.\   267,
195 (1996).

\bibitem{efgo00} J. Ellis, T. Falk, G. Ganis, K. A. Olive,
Phys.\ Rev.\ D62, 075010 (2000).

\bibitem{bgz92} V. Berezinsky, A.V. Gurevich, K.P. Zybin,
Phys.\ Lett.\  B294, 221 (1992).

\bibitem{bbm94} V. Berezinsky,
A. Bottino, G. Mignola,  Phys.\ Lett.\  B325, 136 (1994).

\bibitem{gs99} P. Gondolo, J.  Silk,  Phys.\
Rev.\ Lett.\   83, 1719  (1999).

\bibitem{g00} P. Gondolo,  hep-ph/0002226.

\bibitem{bss01} G. Bertone, G. Sigl, J. Silk,
MNRAS 326, 799 (2001).

\bibitem{begu99} L. Bergstr\"om, J. Edsj\"o, P.
Gondolo, P. Ullio,  Phys.\ Rev.\ D59, 043506  (1999).

\bibitem{cm01}
C. Calc\'{a}neo--Rold\'{a}n, B. Moore, astro-ph/0010056.

\bibitem{abo02} R. Aloisio, P. Blasi, A.V. Olinto, astro-ph/0206036.

\bibitem{bot02} P. Blasi, A.V. Olinto, C. Tyler, astro-ph/0202049.

\bibitem{t02} C. Tyler, astro-ph/0203242

\bibitem{mm02} D. Merritt, M. Milosavljevic, in DARK 2002: 4th
International Heidelberg Conference on Dark Matter in Astro and Particle
Physics, 4-9 Feb 2002, Cape Town, South Africa, H.V.
Klapdor-Kleingrothaus, R. Viollier (eds.)  and astro-ph/0205140.

\bibitem{gmglqs}
S. Ghigna, B. Moore, F. Governato, G. Lake, T. Quinn, J. Stadel,
     MNRAS   300, 146 (1998).

\bibitem{Moore1} B. Moore, S. Ghigna, F. Governato, G. Lake, T. Quinn,
J. Stadel, P. Tozzi, Ap.J. Lett. 524, L19 (1999).

\bibitem{Klypin1} A. Klypin, A.V. Kravtsov, O. Valenzuela, F. Prada,
Ap.J., 522, 82 (1999).

\bibitem{iro02} A. Tasitsiomi, astro-ph/0205464.


\bibitem{veritas} V.V. Vassiliev  et al., 26th ICRC, (Salt Lake City),
5, 299 (1999);  http://veritas.sao.arizona.edu

\bibitem{hess} W. Hofmann, {\it GeV-TeV Gamma Ray Astrophysics Workshop},
Snowbird, UT. (1999); AIP Proc. Conf. 515, 492 (1999).

\bibitem{magic} M. Martinez, 26th ICRC, (Salt Lake City), 5, 219 (1999).

\bibitem{cang3} M. Mori et al., {\it GeV-TeV Gamma Ray Astrophysics
Workshop}, Snowbird, UT.
(1999); AIP Proc. Conf. 515, 485 (1999).

\bibitem{font} A.S. Font, J.F. Navarro, J. Stadel, T. Quinn, Ap.J. 563,
L1 (2001).

\bibitem{waketal} I.R. Walker, J.C. Mihos, L. Hernquist, Ap.J. 460,
121 (1996).

\bibitem{veletal} H. Velazquez, S.D.M. White, MNRAS 304, 254 (1999).

\bibitem{Pasquale1} P. Blasi, R.K. Sheth, Phys.\ Lett.\ B486,
233 (2000).

\bibitem{Lake1} G. Lake, Nature 346, 39 (1990).

\bibitem{Bergstrom2} L. Bergstr\"om, J. Edsj\"o, P. Gondolo,
P. Ullio, Phys.\ Rev.\ D59, 043506 (1999).

\bibitem{navarro1} J.F. Navarro, C.S. Frenk, S.D.M. White, Ap.J. 462,
563 (1996).

\bibitem{klypin} A. Klypin, A.V. Kravtsov, J. Bullock, J.R. Primack,
Ap.J. 554, 903 (2001).


\bibitem{taylor} J.E. Taylor, J.F. Navarro, Ap.J. 563, 483 (2001).

\bibitem{taynav} J.F. Navarro, astro-ph/0110680.

\bibitem{binney} J. Binney, S. Tremaine, {\it Galactic Dynamics},
p. 452 (Princeton:Princeton University Press).

\bibitem{one}
Similar is the conclusion with respect to
the mass, $M_{G}(r_{clump})$, that appears in Eq.(\ref{tidal}).
Namely,  for the distances from the galactic
center that are relevant here, both profiles yield
similar results.

\bibitem{comment} More generally, for an $r^{-\alpha}$ behavior of 
the density profile
at the center, the contribution of the core to the flux scales as 
$R_{core}^{3 -2 \alpha}$, which
implies a sensitive dependence on $R_{core}$ for $\alpha > 1.5$, 
assuming of course
that $R_{core} \ll R_{clump}$. Note however that the behavior of the 
density profile
at intermediate distances ($~r_{s}$) might also be important in some cases
in determining the
overall dependence of the flux on $R_{core}$.

\bibitem{Baltz1} E.A. Baltz, C. Briot, P. Salati, R. Taillet,
J. Silk, Phys.\ Rev.\ D61, 023514 (1999).

\bibitem{Bergstrom3} L. Bergstr\"om, J. Edsj\"o, C. Gunnarsson,
Phys.\ Rev.\ D63, 083515 (2001).

\bibitem{Edsjo1} J. Edsj\"o, P. Gondolo, Phys. Rev. D56,
01879 (1997).

\bibitem{darksusy1} DarkSUSY, http://www.physto.se/$\sim$edsjo/dark susy,
P.Gondolo, J. Edsj\"o, L.Bergstr\"om, P.Ullio, E.A. Baltz, in preparation.


\bibitem{Bergstrom4} L. Bergstr\"om, P. Gondolo, Astrop.Phys.
5, 263  (1996).

\bibitem{Bergstrom8} L. Bergstr\"om, P. Ullio, Nucl. Phys. B504, 27 (1997).

\bibitem{Bergstrom9} P. Ullio, L. Bergstr\"om,
Phys.\ Rev.\ D57, 1962 (1998).

\bibitem{hfast1} M. Drees, M.M. Nojiri, D.P. Roy, Y.Yamada,
Phys.\ Rev.\ D56, 276 (1997).

\bibitem{hfast} S. Heinemeyer, W. Hollik, G. Weiglein,
Phys. Lett. B455, 179 (1999); http://www-itp.physik.uni-
karlsruhe.de/feynhiggs/

\bibitem{Gondolo1} P. Gondolo, G. Gelmini, Nucl. \ Phys.\  B360, 145
(1991).

\bibitem{Peacock1} J.A. Peacock, {\it Cosmological Physics},
section 3.2 (Cambridge: Cambridge University Press).

\bibitem{reiss1} A.G. Reiss et al., A.J. 116, 1009 (1998).

\bibitem{Perlmutter1} S. Perlmutter et al., Ap.J.  517, 565 (1999).

\bibitem{hill1} C.T. Hill , Nucl.\ Phys.\ B224, 469 (1983).

\bibitem{hill2}  C.T. Hill, D.N. Schramm, T.P. Walker,
Phys. \ Rev.\ D36, 1007 (1987).

\bibitem{two} For simplicity, from that point on
we will use the term cross section to denote
   the thermally averaged
product of the cross section times the relative velocity.

\bibitem{unre}
In fact, the  low (good) VERITAS
resolution -- which renders the  nearest clumps resolved -- has
been a crucial factor in determining our strategy, namely, the use of 
individual clumps rather than of all-sky maps that might be more 
appropriate in the case of numerous, unresolved clumps, as most 
clumps would appear when observed via instruments of higher (worse) 
angular resolution, e.g., GLAST.

\bibitem{Bergstrom6} L. Bergstr\"om, P. Ullio, J.H. Buckley,
Astrop.Phys.  9, 137 (1998 ).

\bibitem{Longair}  M.S. Longair, {\it High Energy Astrophysics}
(Cambridge: Cambridge University Press).

\bibitem{Sreekumar} P. Sreekumar et al., Ap.J. 494, 523 (1998).

\bibitem{Hunter} S.D. Hunter et al., Ap.J. 481, 205 (1997).

\bibitem{Bergstrom7} L. Bergstr\"om, J. Edsj\"o, C. Gunnarsson,
Phys.\ Rev.\ D63, 083515 (2001).

\bibitem{glast}
D.A. Kniffen, D.L. Bertsch, N. Gehrels,
{\it GeV-TeV Gamma Ray
Astrophysics Workshop}, Snowbird, UT. AIP Proc. Conf.
515,  492 (1999).

\bibitem{three} Other weaker dependencies exist as, for example, via
Eqs.(\ref{joint}) and (\ref{distance}).

\end{thebibliography}
\end{document}